\documentclass[12pt,a4paper]{article}

\usepackage{latexsym, amsfonts,amsmath,amssymb,graphics,bbm}
\usepackage{pifont}    
\usepackage{dsfont}
\usepackage{feynmp}    
\usepackage{nccmath}   
\usepackage[dvips]{graphicx}

\usepackage{mathrsfs}  
\usepackage{bbm} 
\usepackage{slashed}   
\DeclareMathAlphabet{\mathpzc}{OT1}{pzc}{m}{it}

\newcommand{\vphi}{{\pmb{\phi}}}
\newcommand{\rmt}{{\scriptscriptstyle\top}} 

\newcommand{\kA}[0]{}
\newcommand{\kB}[0]{}
\newcommand{\kC}[0]{}
\newcommand{\kD}[0]{}

\newcommand{\kF}[0]{}
\newcommand{\kG}[0]{}
\newcommand{\kH}[0]{}
\newcommand{\kI}[0]{}
\newcommand{\kJ}[0]{}
\newcommand{\kK}[0]{}
\newcommand{\kL}[0]{}

\newcommand{\kN}[0]{}

\newcommand{\kR}[0]{}
\newcommand{\kS}[0]{}
\newcommand{\kT}[0]{}
\newcommand{\kU}[0]{}

\newcommand{\kW}[0]{}
\newcommand{\kX}[0]{}
\newcommand{\kY}[0]{}
\newcommand{\kZ}[0]{}

\newcommand{\kBB}[0]{}
\newcommand{\kCC}[0]{}
\newcommand{\kDD}[0]{}
\newcommand{\kEE}[0]{}
\newcommand{\kFF}[0]{}
\newcommand{\kGG}[0]{}
\newcommand{\kHH}[0]{}
\newcommand{\kII}[0]{}
\newcommand{\kJJ}[0]{}
\newcommand{\kKK}[0]{}
\newcommand{\kLL}[0]{}
\newcommand{\kMM}[0]{}
\newcommand{\kNN}[0]{}
\newcommand{\kOO}[0]{}
\newcommand{\kPP}[0]{}
\newcommand{\kQQ}[0]{}
\newcommand{\kRR}[0]{}
\newcommand{\kSS}[0]{}
\newcommand{\kTT}[0]{}
\newcommand{\kUU}[0]{}
\newcommand{\kVV}[0]{}
\newcommand{\kWW}[0]{}
\newcommand{\kXX}[0]{}
\newcommand{\kYY}[0]{}
\newcommand{\kZZ}[0]{}

\newcommand{\kAAA}[0]{}
\newcommand{\kBBB}[0]{}
\newcommand{\kCCC}[0]{}
\newcommand{\kDDD}[0]{}
\newcommand{\kEEE}[0]{}
\newcommand{\kFFF}[0]{}
\newcommand{\kGGG}[0]{}

\newcommand{\kJJJ}[0]{}
\newcommand{\kKKK}[0]{}

\newcommand{\kMMM}[0]{}
\newcommand{\kNNN}[0]{}
\newcommand{\kOOO}[0]{}
\newcommand{\kPPP}[0]{}
\newcommand{\kQQQ}[0]{}
\newcommand{\kRRR}[0]{}
\newcommand{\kSSS}[0]{}

\newcommand{\kUUU}[0]{}

\newcommand{\kYYY}[0]{}
\newcommand{\kZZZ}[0]{}

\newcommand{\koment}[1]{}      
\newcommand{\nothing}[1]{}    

\renewcommand{\a}[0]{{(a)}}

\newcommand{\ds}[1]{\slashed{#1}}

\newcommand{\anti}[1]{\overline{#1}}

\newcommand{\derf}[2]{\frac{\delta #1}{\delta #2}}     

\newcommand{\dd}[0]{{\rm d}}

\newcommand{\re}[0]{{\rm Re}}
\newcommand{\im}[0]{{\rm Im}}

\newcommand{\diag}{{\rm diag}}
\newcommand{\hb}{\hbar}
\newcommand{\id}[0]{\mathds{1}}

\newcommand{\refer}[1]{(\ref{#1})}

\newcommand{\nn}[0]{\nonumber}
\newcommand{\tm}[0]{{\widetilde{m}}}

\newcommand{\tP}[0]{{\widetilde{P}}}

\newcommand{\cC}[0]{{\mathcal{C}}}

\newcommand{\cF}[0]{{\mathcal{F}}}

\newcommand{\cJ}[0]{{\mathcal{J}}}

\newcommand{\cL}[0]{{\mathcal{L}}}

\newcommand{\cN}[0]{{\mathcal{N}}}
\newcommand{\cO}[0]{{\mathcal{O}}}

\newcommand{\cR}[0]{{\mathcal{R}}}

\newcommand{\cU}[0]{{\mathcal{U}}}
\newcommand{\cV}[0]{{\mathcal{V}}}

\newcommand{\cY}[0]{{\mathcal{Y}}}

\newcommand{\cc}[0]{{\mathpzc{c}}}

\newcommand{\cf}[0]{{\mathpzc{f}}}
\newcommand{\cg}[0]{{\mathpzc{g}}}

\newcommand{\cu}[0]{{\mathpzc{u}}}

\newcommand{\sD}[0]{{\mathscr{D}}}

\newcommand{\sG}[0]{{\mathscr{G}}}

\newcommand{\sI}[0]{{\mathscr{I}}}

\newcommand{\sM}[0]{{\mathscr{M}}}

\newcommand{\sU}[0]{{\mathscr{U}}}

\newcommand{\sX}[0]{{\mathscr{X}}}

\newcommand{\sZ}[0]{{\mathscr{Z}}}

\newcommand{\bC}[0]{{\mathbb{C}}}

\newcommand{\bM}[0]{{\mathbb{M}}}

\newcommand{\bP}[0]{{\mathbb{P}}}

\newcommand{\bR}[0]{{\mathbb{R}}}
\newcommand{\bS}[0]{{\mathbb{S}}}

\newcommand{\bY}[0]{{\mathbb{Y}}}

\newcommand{\al}{\alpha}
\newcommand{\be}{\beta}
\newcommand{\de}{\delta}
\newcommand{\De}{\Delta}
\newcommand{\ga}{\gamma}
\newcommand{\Ga}{\Gamma}
\newcommand{\si}{\sigma}
\newcommand{\Si}{\Sigma}

\newcommand{\la}{\lambda}
\newcommand{\La}{\Lambda}
\def\th{\theta} 
\newcommand{\Th}{\Theta}

\newcommand{\vep}{\varepsilon}
\newcommand{\om}{\omega}
\newcommand{\Om}{\Omega}
\newcommand{\ze}{\zeta}

\newcommand{\YF}[0]{Y}     

\newcommand{\TS}[0]{\mathcal{T}}       
\newcommand{\TF}[0]{{ \mathfrak{f}}}

\newcommand{\volel}[1]{\!\! {{\rm d}^4 #1 }}

\newcommand{\volfour}[1]{\!\! \frac{{\rm d}^4 #1}{\left(2\pi\right)^4}}

\newcommand{\VEV}[1]{\left<#1\right>}

\begin{document}
\koment{{{\bf  !!! COMMENT !!! }}}

\pagestyle{empty}

\begin{center}

{\large {\bf LSZ-reduction, resonances and non-diagonal propagators: 
fermions and scalars}}
\vspace*{1cm}

{Adrian Lewandowski 
\footnote{Present address:\\
\emph{Albert Einstein Center for Fundamental Physics,\\ 
Institute for Theoretical Physics, University of Bern, \\
Sidlerstrasse 5, CH-3012 Bern, Switzerland\\ }}$^{,}$\footnote{Email: lewandowski.a.j@gmail.com}
}

\vspace{1cm}
{{\it
Max-Planck-Institut f\"ur Gravitationsphysik
(Albert-Einstein-Institut)\\
Am M\"uhlenberg 1, D-14476 Potsdam, Germany
}}

\vspace*{1.7cm}
{\bf Abstract}
\end{center}
\vspace*{5mm}
\noindent

We analyze in details the effects associated with mixing of fermionic fields. 
In a system with an arbitrary number of Majorana or Dirac 
particles, 
a simple proof of factorizability of residues of non-diagonal 
propagators at the complex poles is given, together with a prescription 
for finding the ``square-rooted" residues to all orders of perturbation 
theory, in an arbitrary renormalization scheme.  
Corresponding prescription for the scalar case is provided as well.  

\vspace*{1.0cm}


\vspace*{0.2cm}

\vfill\eject
\newpage

\pagestyle{plain}

\tableofcontents

\section{Introduction}\label{Sec:Intro}

Finding the correct form of renormalization conditions in gauge theories of 
particle physics is a nontrivial task, as the improper conditions can 
spoil the gauge symmetry even in the presence of a symmetry preserving 
regularization. 
For this reason, renormalization schemes with implicit renormalization 
conditions are often employed. A famous representative of this class is the 
$\overline{{\rm MS}}$ scheme of dimensional regularization \cite{MSbar}. 
Zinn-Justin proved 
that non-Abelian gauge symmetries are preserved 
in minimal subtraction schemes based on a symmetry preserving regularization 
\cite{Z-J:74}, what makes $\overline{{\rm MS}}$ the most 
convenient choice, at least in non-chiral theories. In fact, the 
violation 
of gauge symmetry induced by regularization can be handled according to the 
general rules of ``algebraic renormalization" \cite{BRS1,BRS2,PiguetSorella} (see also \cite{Bon}  
for a discussion in the context of dimensional regularization with the consistent 't~Hooft-Veltman-Breitenlohner-Maison prescription for $\gamma^5$).  
This approach does not preclude the use of schemes with implicit 
renormalization conditions. 
For instance, 
in \cite{ChLM} a renormalization scheme for gauge 
theories in the cutoff regularization was constructed by carefully selecting a 
set of vertices to which  
non-minimal (cutoff-independent) counterterms are added. 
These non-minimal counterterms are completely fixed by investigating the 
Slavnov-Taylor identities for the gauge symmetry, what leads to a mass 
independent renormalization scheme with implicit renormalization 
conditions. In such a framework, renormalization can be 
considered as the first stage in the study of correlation functions, a stage 
which is 
independent of the physical interpretation of the studied model. 

In this paper we are interested \emph{only} in the second stage, i.e. in the 
extraction 
of scattering amplitudes from \emph{renormalized} 
correlation functions in an arbitrary 
renormalization scheme.\footnote{Of course, 
it is assumed that the scheme preserves Slavnov-Taylor identities of the gauge 
symmetry.} 
More 
specifically, we are interested in amplitudes corresponding to the particles 
described by the system of mixed fields with non-diagonal propagators. 
In this 
context, the crucial objects are residues of the (renormalized) 
propagators; for 
instance in the case of scalar fields one finds the following expression for the 
full propagator 
\eq{\label{Eq:PolePart} 
\langle T(\phi^k(x)\phi^j(y))\rangle=
\int\volfour{p}\,\, e^{-i\,p(x-y)}
\bigg\{\!\sum_\ell 
\, 
\ze^k_{S[\ell]} 
\,
\frac{i}{p^2-m^2_{S(\ell)}}
\,\ze^j_{S[\ell]}
+
\left[\!\!\begin{array}{c}
\text{non-pole}\\ \text{part}
\end{array}\!\!\right]
\!\bigg\}
\,.
}
The factorization of the residues at the (complex) poles is a well-known 
property \cite{Stapp}.  For the mixed $Z$-photon system of the Standard 
Model it was explicitly demonstrated in \cite{Stuart}. For fermionic systems 
the factorization was shown in 
\cite{PilafRL1,PlumOld} (see also \cite{PilafUnd,PlumNew}). 
\footnote{We also mention in this context Ref. \cite{AOKI} where the 
factorization of the residues was shown for scalar, vector and fermionic systems, on the basis of the presupposed existence of the on-shell 
renormalization scheme. This approach makes the generalization to the complex 
poles rather difficult.} 
   
The coefficients $\ze^k_{S[\ell]}$ associated with real poles $m^2_{S(\ell)}$ are crucial for obtaining correctly normalized (i.e. consistent with unitarity) 
transition amplitudes between the states of stable particles.  
On account of Cutkosky-Veltman rules, it is well-known  \cite{Velt}  that the 
$S$-matrix  
is unitary provided that (1.) unstable particles appear only as internal lines, 
and (2.) asymptotic (free) fields appearing in the LSZ-reduced formula for the 
$S$-operator (see e.g. \cite{Bec} and Eq. \refer{Eq:S} below) are normalized so 
as to reproduce the behavior of full propagators about the poles associated with 
stable particles. Thus, the asymptotic field $\vphi^j$ associated with $\phi^j$ can be written as 
$\vphi^j=\sum{}^{\prime}_{\ell} 
{\zeta}^j_{S[\ell]}  
\Phi^{\ell}$,
where $\Phi^{\ell}$ is a canonically normalized  
free scalar field of mass $m_{S(\ell)}$, and the summation runs over indices 
labeling real poles. 

For resonances the external lines, aka the 
``in" and ``out" states, do not really exist.  Nonetheless, 
coefficients $\ze^k_{S[\ell]}$ associated with the complex poles 
are useful in studying properties of unstable particles 
\cite{Stapp, Stuart, PilafRL1, PlumOld}. In this connection, the problem of 
finding a convenient prescription for coefficients $\ze^k_{S[\ell]}$ 
parameterizing the residues in Eq. \refer{Eq:PolePart}  
have gained renewed interest in recent years.  
In Ref. \cite{Fuchs} the case of 3-by-3 mixing of scalar particles was analyzed 
in details. The factorization property \refer{Eq:PolePart} was demonstrated and 
explicit formulae for the coefficients $\ze^k_{S[\ell]}$ were given. 
These results were applied to the neutral Higgs sector of MSSM; it was shown that cross-sections 
obtained by neglecting the non-pole part in Eq. \refer{Eq:PolePart} agree to 
good accuracy with the cross-sections based on full propagators.  
Analysis of a generic $n$-by-$n$ mixing in fermionic systems was given in 
\cite{Kniehl-1,Kniehl-2,Kniehl-3}. 

The purpose of this paper is to generalize and to simplify the available in the 
literature procedure of calculating coefficients $\ze$  for 
fermions and scalars. Our analysis is closest in spirit to the one given in 
\cite{Kniehl-1,Kniehl-2,Kniehl-3}; there are, however, some differences. 
First, we follow the philosophy of keeping the renormalization scheme as general 
as possible. In particular, we do not impose any concrete renormalization 
conditions on the two-point functions. Second, we offer a technical improvement in comparison with the analyzes of 
\cite{Stapp,Kniehl-1,Kniehl-2,Kniehl-3}, where the cofactor matrix 
was used to get the formulae for $\ze$. 
By contrast, the coefficients $\ze$ in our approach are expressed directly 
in terms of properly normalized eigenvectors of certain ``mass-squared 
matrices", so that the case of degenerated eigenvalues is naturally covered by 
our prescription. Thus, the proposed prescription for finding $\ze$ can be considered as a  
generalization of the standard 
procedure for finding tree-level mass eigenstates.   
The analysis of mixed vector fields along these lines will be given elsewhere \cite{OnFeRu2}.

The formulae presented below are valid in the generic case of mixed unstable Majorana or/and Dirac fermions (a scalar version is also given). For completeness, the expression for the Landau-gauge one-loop fermionic self-energy of a general renormalizable model in the $\anti{{\rm MS}}$ scheme is provided below. 
The paper is therefore intended to cultivate a long tradition by providing generic formulae that can be easily applied to (almost) any model at hand, especially in the present computer era, see e.g. 
\cite{Weinberg}, \cite{Gross-Wilczek}, \cite{Jack-Osborn}, 
\cite{Mach-Vaughn}, \cite{Martin}, \cite{Martin-3}, \cite{Luo}, 
\cite{ChLM}.       

The remainder is organized as follows. In the next section the notation is specified, together with basic assumptions. In Sec. \ref{Sec:PrescrFerm} 
the prescription for $\zeta$ matrices is given for massive Majorana particles 
(\ref{Sec:PrescrFerm}.A), massive Dirac particles (\ref{Sec:PrescrFerm}.B), generic spin-1/2 fermions (\ref{Sec:PrescrFerm}.C) and scalars (\ref{Sec:PrescrFerm}.D),  together with the generic expression for fermionic one-loop self-energy (\ref{Sec:PrescrFerm}.E). The correctness of the prescription is proved in Sec. \ref{Sec:Proof} and the last section is reserved for conclusions.

\kZZZ 

\section{Notation and assumptions}\label{Sec:NotAndAssum}
\noindent In most formulae given below indices are suppressed and matrix multiplication is understood. The summation convention is used only when an upper index is contracted with a lower one; whenever ambiguities may arise, sums are explicitly present. The Minkowski metric has the form 
\eq{
\eta=[\eta_{\mu\nu}]=\diag(+1,-1,-1,-1)\,.
}  

Recall that a Majorana field \cite{WeinT3} $\psi^{\widetilde{a}}$ is a pair of a Weyl field $\chi^{a}_A$, below referred to as the left-chiral Weyl field (LW), 
and its Weyl conjugate $\anti{\chi}^{a\dot{A}}$, alias the right-chiral Weyl field  (RW) 
\eq{\label{Eq:MajoField}
\psi^{\widetilde{a}}
=
\left[
\begin{array}{l}
\chi^{a}_A\\
\anti{\chi}^{a\dot{A}}\\
\end{array}
\right]
,
}
here $a=1,\ldots,n$ is a generalized-flavor index, $A$ and $\dot{A}$ are $SL(2,\bC)$ indices, while $\widetilde{a}=(a,(A,\dot{A}))$. 

Take, for instance, (a toy version of) the Standard Model \cite{WeinT2} in which all Weyl fields except for these that describe the electron-positron pair of states have been forgotten. Let $\la_A$ be a LW representing the charged component of the lepton (would-be) $SU(2)_L$-doublet and let $\rho_A$ be a LW of the charged lepton $SU(2)_L$-singlet. In this case $n=2$ and the fields with the definite generalized-flavor (henceforth called flavor) can be chosen as 
\eq{\label{Eq:PhysBas}
\chi^{1}=\la\,,\qquad\qquad  
\chi^{2}=\rho, 
}
though nothing (but common sense) prevents a more general choice 
\eq{\label{Eq:cu}
\chi^{a} = \cu^{a}_{\ \,1}\, \lambda + \cu^{a}_{\ \, 2}\, \rho\,,
} 
with an arbitrary unitary matrix $\cu$, which off-diagonalizes the charge generator.  
\\

With chiral projections $P_{L,R}$ 
\eq{\nn
P_L\,\psi\simeq \left[
\begin{array}{l}
\chi\\
0\\
\end{array}
\right]
\,,\qquad
P_R\,\psi\simeq \left[
\begin{array}{l}
0\\
\anti{\chi}\\
\end{array}
\right]
\,,
}
and the charge conjugation matrix $\cC$ that expresses the Dirac conjugate of $\psi$
 in terms of $\psi$ itself 
\eq{\nn 
\bar{\psi}
=
\psi^{\rmt} \cC\,,
}
the renormalized (in some renormalization scheme) one-particle-irreducible (1PI) two-point function of Majorana fileds can be written in the following form 
\koment{str.7} 
\eqs{\label{Eq:Gamma2} 
\widetilde{\Ga}_{\widetilde{a}\widetilde{b}}(-p,p)
&=&
\bigg[
\cC\Big\{
\phantom{+}\!\!\!\!
\left(\ds{p}\sZ_L(p^2)-\sM_L(p^2)\right)P_L
+\qquad\nn\\
&{}&\,\,\,\,\,\,\,\,+\!
\left(\ds{p}\sZ_R(p^2)-\sM_R(p^2)\right)P_R
\Big\}
\bigg]_{\widetilde{a}\widetilde{b}}
\,\, ,
\qquad
}
where
\eq{\label{Eq:Z_LR-expansion}
\sZ_{L,R}(p^2)=\mathds{1}+\cO(\hbar)\,.
}
Clearly, matrices $P_{R,L}$, $\cC$ and $\ds{p}$ carry only the $SL(2,\bC)$ indices, 
while $\sZ_{L,R}$ and $\sM_{L,R}$ carry only the flavor indices; the tensor products $\otimes$ are not explicitly shown in Eq. \refer{Eq:Gamma2}. 
($\sZ_{L,R}$ and $\sM_{L,R}$ are, essentially, the 1PI functions of different pairs of Weyl fields; Majorana fields are used here and below only for bookkeeping reasons.)

The full propagator of Majorana fields is given by 
\koment{str.3}
\eq{\label{Eq:G2} 
\widetilde{G}^{\,\widetilde{a}\widetilde{b}}(p,-p)
=
i\Big[\widetilde{\Ga}(-p,p)^{-1}\Big]^{\widetilde{a}\widetilde{b}}
=
i
\Big[ \hat{\sD}_{\cF}(p)\, \cC^{-1} \Big]^{\widetilde{a}\widetilde{b}}
\,.\
}
Inverting the two-point function in Eq. \refer{Eq:Gamma2} one finds
\koment{str.8}
\eq{\label{Eq:D-L+R}
\hat{\sD}_{\cF}(p)
=
P_L\,\hat{\sD}_{L}(p)+P_R\,\hat{\sD}_{R}(p),
}
where ($s\equiv p^2$) \koment{str.Z2}
\eqs{\label{Eq:D-LR}
\nn
\hat{\sD}_{L}(p)&=&
\left[s\mathds{1}-\bM^2_L(s)\right]^{-1}\!
\sZ_L(s)^{-1}\left[\ds{p}+\sM_R(s)\sZ_R(s)^{-1}\right],
\\ \nn
\hat{\sD}_{R}(p)&=&
\left[s\mathds{1}-\bM^2_R(s)\right]^{-1}\!
\sZ_R(s)^{-1}\left[\ds{p}+\sM_L(s)\sZ_L(s)^{-1}\right],
\\ &{}&
}
and \koment{str.Z1} 
\eqs{\label{Eq:bM2-LR}
\nn
\bM^2_L(s)
&=&
\sZ_L(s)^{-1}\,\sM_R(s)\,\sZ_R(s)^{-1}\,\sM_L(s)\,,
\\
\bM^2_R(s)
&=&
\sZ_R(s)^{-1}\,\sM_L(s)\,\sZ_L(s)^{-1}\,\sM_R(s)\,.
}
The antisymmetry of the fermionic two-point function, Eq. \refer{Eq:Gamma2}, yields 
\koment{str.12}
\eqs{\label{Eq:AntiSym}
\nn
\sM_X(s)&=&\sM_X(s)^{\rmt}\,, \qquad X=L,R, \\
\sZ_R(s)&=&\sZ_L(s)^{\rmt}\,,
}
and thus
\koment{str.Z2}
\eq{\label{Eq:bMRtrans}
\bM^2_R(s)^\rmt
=
\sZ_L(s)\,\bM^2_L(s)\,\sZ_L(s)^{-1}\,,
}
what gives
\koment{str.32} \kA

\eq{\label{Eq:sX}
\sX(s)\equiv\det(s\mathds{1}-\bM^2_L(s))
=
\det(s\mathds{1}-\bM^2_R(s))\,,
} 
hence the poles of both chiral parts $\hat{\sD}_{L,R}$ of propagator in Eq. \refer{Eq:D-L+R}  appear at the same points $s=m_{(a)}^2$, obeying the condition \kOOO
\eq{\label{Eq:GapEq-0}
\sX(m_{(a)}^2)=0\,.
}

In this paper three technical assumptions are made about the solutions to Eq. \refer{Eq:GapEq-0} and the matrices $\bM^2_L(m_{(a)}^2)$. First, it is assumed that each generalized eigenvector \kFF (see e.g. \cite{Axler})  of $\bM^2_L(m_{(a)}^2)$ associated with the eigenvalue $m_{(a)}^2$ is an (ordinary) eigenvector; in other words, it is assumed that in the Jordan basis for $\bM^2_L(m_{(a)}^2)$ the block corresponding to $m_{(a)}^2$ is diagonal. This excludes standard pathologies associated with non-diagonalizable matrices (e.g. second order poles of gauge-field propagators in covariant non-Feynman gauges caused by {\emph{pseudo}}Hermiticity of the Hamiltonian \cite{NAKANISHI}). \kB 

Second, it is assumed that each nonzero solution $m_{(a)}^2$ is nonzero  at the tree level, as is usually the case in the common seesaw models. 

Third, it is assumed that, roughly speaking, the quantum corrections do not change the total number of solutions to Eq. \refer{Eq:GapEq-0}. More specifically, \kAAA 
suppose that the $a$ label distinguishes different solutions $m_{(a)}^2$. Let 
$\xi_{[a_1]},\, \xi_{[a_2]}\,, \ldots\,, $ be a basis of the eigenspace of $\bM^2_L(m_{(a)}^2)$ associated with the eigenvalue $m_{(a)}^2$. It is assumed that each element in the sequence 
\eq{\nn 
\xi_{[1_1]},\,\ldots\,, \xi_{[2_1]}\,\,\ldots\,,
}
has the form $\xi_{[a_r]} = \xi_{[a_r]}^0 + \cO(\hbar)$, where vectors
\eq{\nn 
\xi_{[1_1]}^0,\,\ldots\,, \xi^0_{[2_1]}\,\,\ldots\,,
}
are of zeroth order in $\hbar$ and form a basis of $\bC^n$, with $n$ denoting the total number of LWs. \footnote{The reader should be warned that the $a$ label on pole masses is the same as the index on flavor eigenfields $\chi^a$, even though $\chi^a$ are not assumed to be the eigenstates of the tree-level (nor the pole) masses. This little abuse of notation will not lead to any misunderstandings.} \kMM \kEE

The pole masses $m_{(a)}^2$ are formal power series in $\hb$. Thus, if all of the tree-level masses of fermions are different, then $\bM^2_L(s)$ is diagonalizable as a formal power series 
\eq{\nn
\bM_L^2(s)=W(s)^{-1}\,\diag\big(d_1(s),...,d_n(s)\big)\, W(s)\,,
}
and 
\eq{\nn
\sX(s)=\prod_{p=1}^{n}\Big(s-d_{p}(s)\Big)\,.
}
If $d_{a}(s)=(m^{\rm tree}_{(a)})^2+\cO(\hb)$, then Eq. \refer{Eq:GapEq-0} reads
\eq{\nn
d_{a}(m_{(a)}^2)=m_{(a)}^2\,,
}
and has a unique solution \kGG 
$m_{(a)}^2=(m^{\rm tree}_{(a)})^2+\cO(\hb)$. In particular, the first and the third assumption are satisfied in this case. \kL 
In general, assuming non-degeneracy of the tree-level masses is however not an option as physics is about symmetries. Therefore it is convenient (and desirable from practical point of view) to distinguish two special situations called below the Majorana case and the Dirac case.

Let $\sG$ be the group 
of exact, linearly realized, internal global symmetries \kJJ of the tree-level action that are respected by the renormalization conditions and let $\cU(\cdot)$ be the representation of $\sG$ on the left-chiral flavor eigenfields $\chi^a$. The two-point function \refer{Eq:Gamma2} obeys the following conditions \koment{str.Z.53}
\eqs{\label{Eq:SymIntCond}
\nn
\sM_L(s) &=& \cU(\cg)^\rmt\sM_L(s)\,\cU(\cg) \,,\\
\nn
\sM_R(s)  &=& \cU(\cg)^\dagger\sM_R(s)\,\cU(\cg)^\star\,,\\
\nn
 \sZ_L(s)  &=&  \cU(\cg)^\dagger \sZ_L(s)\,\cU(\cg)\,,\\
 \sZ_R(s) &=& \cU(\cg)^\rmt \sZ_R(s)\,\cU(\cg)^\star\,, \qquad \forall_{\cg\in\sG}\,.
}

Consider first a toy model in which fermions form three families, each one consisting of gluinos of the Minimal Supersymmetric Standard Model \cite{WeinT3}. In this case the flavor index is a pair of an adjoint color index and a family index, and the most general matrices $\sZ_{L,R}$ and  $\sM_{L,R}$ consistent with the $SU(3)_C$ symmetry have the form \koment{str.$O^3$}
\eqs{\label{Eq:Maj-deg}
\nn
\sM_{L,R}(s)&=&\mathds{1}_{k\times k} \otimes \sM^{\text{fam}}_{L,R}(s)\,,\\
\sZ_{L,R}(s)&=&\mathds{1}_{k\times k} \otimes \sZ^{\text{fam}}_{L,R}(s)\,,
}
and thus
\eqs{ \label{Eq:Maj-deg-bMkw}
\ \  \ 
\bM^2_{L}(s)&=&\mathds{1}_{k\times k} \otimes \bM^{2\,\text{fam}}_{L}(s)\,,
}
where $\mathds{1}_{k\times k}$ (with $k=8$) is the identity matrix in the adjoint color space, while $\sM^{\text{fam}}_{L,R}$ and $\sZ^{\text{fam}}_{L,R}$  are $3\times 3$ matrices in the family space. 
In particular,  $\bM^2_L(m_{(a)}^2)$ are diagonalizable if e.g. the tree-level contribution to $\bM^{2\,\text{fam}}_L(0)$ has non-degenerate eigenvalues. 
A situation in which the two-point functions have the form \refer{Eq:Maj-deg} with an arbitrary number $f$ of ``families", an arbitrary $k$, and with $f$ different and nonvanishing eigenvalues of the tree-level contribution to $\bM^{2\,\text{fam}}_L(0)$ is called below the Majorana case; the total number of flavors equals $n=f\times k$. As far as the propagator and mixing are concerned, one can in this case restrict attention to a single color. %
\footnote{
A more physical representative of the Majorana case is the type I seesaw mechanism with $k=1$ and $f=3+3$ neutrinos.} It is worth emphasizing that 
the Majorana case (as well as the Dirac case below) is defined here by demanding $m_\a^{\rm tree}\neq0$ for all $a$, in order to make the corresponding prescription in Sec.  \ref{Sec:PrescrFerm}.A (respectively, \ref{Sec:PrescrFerm}.B) as simple and practical as possible.\footnote{
\mbox{In light} of neutrino oscillations, theories with massless spin-$1/2$ fermions are no longer so appealing. In fact even in the \emph{pure} SM, symmetries exclude not only neutrino masses but also any mixing between, say, the muon-neutrino and other fermions, and thus allow to restrict the attention to the block of massive fermions.
}
Vanishing masses require a separate treatment and they are dealt with in Sec. \ref{Sec:PrescrFerm}.C devoted to the generic case. 

Consider next a more interesting example of three families of down-type quarks \kCC in the SM (clearly, the $SU(3)_C\times U(1)_Q$ symmetry of the SM prohibits down-type quarks from mixing with other SM fermions). 
Without loss of generality, it can be assumed that the flavor eigenfields $\chi^{a}$ have been chosen so that the 
anti-Hermitian generator of $U(1)_Q$ is diagonal \kC
\eqs{
\cf_Q  &=& {\mathds{1}}_{\ell\times \ell} \otimes
\left[
\begin{array}{cc}
 -\frac{i\,e}{3}\mathds{1}_{3\times 3} & 0 \\
 0 &  \frac{i\,e}{3}\mathds{1}_{3\times 3} \\
\end{array}
\right]\,,\nn\\
&{}&\hspace*{-22 pt}
\cU(\cg^Q_t)=\exp(t\, \cf_Q)\,,\nn
}
where $\mathds{1}_{\ell\times \ell}$ (with $\ell=3$) is the identity matrix in the color space. 
The most general matrices $\sZ_{L,R}$ and  $\sM_{L,R}$ consistent with Eqs. \refer{Eq:AntiSym} and the $SU(3)_C\times U(1)_Q $ symmetry read \koment{str.$A^3$} 
\eqs{\label{Eq:Dirac-Case}
\sZ_L(s)&=&\sZ_R(s)^{\rmt}
= {\mathds{1}}_{\ell\times \ell} \otimes
\left[
\begin{array}{cc}
 \sI_{+}(s)^{-1}  & 0  \\
 0 & \sI_{-}(s)^{-1}  \\
\end{array}
\right]\,,
\nn\\
\sM_X(s)&=&
 {\mathds{1}}_{\ell\times \ell} \otimes
\left[
\begin{array}{cc}
 0     & \mu_X(s) \\
 \mu_X(s)^\rmt &  0 \\
\end{array}
\right]\,, \quad X=L,R,
\nn\\
}
where $\mu_{L,R}$ and $\sI_{\pm}(s)$ are arbitrary $3\times 3$ matrices and, in addition, $\sI_{\pm}(s)$ are nonsingular. Thus \koment{str.$S^3$}
\eq{\label{Eq:Dirac-Case-bMLkw}
\bM^2_L(s)=
{\mathds{1}}_{\ell\times \ell} \otimes
\left[
\begin{array}{cc}
 \bM^2_{+}(s)  & 0  \\
 0 & \bM^2_{-}(s)  \\
\end{array}
\right]\,,
}
with 
\eqs{\label{Eq:Dirac-Case-bMpmkw}
\bM^2_{+}(s)&=&\sI_{+}(s)\mu_{R}(s)\sI_{-}(s)^{\rmt}\mu_{L}(s)^{\rmt}\,,
\nn\\
\bM^2_{-}(s)&=&\sI_{-}(s)\mu_{R}(s)^\rmt\sI_{+}(s)^{\rmt}\mu_{L}(s)\,.
}
Using the relation (valid if, e.g. tree-level masses are non-vanishing, so that $\mu_L(s)$ is nonsingular)
\eq{
\bM^2_{-}(s)^{\rmt} = 
\mu_L(s)^{\rmt}\, \bM^2_{+}(s) \left\{\mu_L(s)^{\rmt} \right\}^{-1}\,,
}
one gets 
\eq{\label{Eq:+Eq-}
\sX_{+}(s)\equiv\det(s\mathds{1}-\bM^2_{+}(s))
=
\det(s\mathds{1}-\bM^2_{-}(s))\,,
} 
hence the determinant in  Eq. \refer{Eq:sX} reads
\eq{\label{Eq:sX-Dirac}
\sX(s)=\sX_{+}(s)^{2\ell}\,.
} 
It follows from Eq. \refer{Eq:+Eq-} that complex poles corresponding to the left-chiral flavor eigenfields with opposite charges are located at the same points $s=m_{(a)}^2$. \kD  A situation in which the two-point functions have the form \refer{Eq:Dirac-Case} with an arbitrary number $f$ of families, an arbitrary $\ell$, and with $f$ different and nonvanishing eigenvalues of the tree-level contribution to $\bM^{2}_+(0)$ is called below the Dirac case (the total number of flavors equals $n=2\times f\times \ell$). 
Once again, as far as the propagator and mixing are concerned, one can in this case restrict attention to a single color, i.e. one can effectively neglect color factors  ${\mathds{1}}_{\ell\times \ell}$ in Eqs. 
\refer{Eq:Dirac-Case}-\refer{Eq:Dirac-Case-bMLkw}.

A simple prescription for the pole part of the propagator \refer{Eq:D-L+R} is given in the next section for these two special cases. A generalization to arbitrary $\bM^2_L(m_{(a)}^2)$ matrices consistent with three assumptions stated above is provided as well. 

It should be noted, however, that infrared problems (see e.g. \cite{Kibble}) are \emph{not} discussed in this paper. In other words,  it is assumed that an IR regulator was introduced (if necessary) so that the propagators do have simple poles at the points obeying Eq. \refer{Eq:GapEq-0}. \kUUU

\section{Prescription}\label{Sec:PrescrFerm}

\noindent {\bf \ref{Sec:PrescrFerm}.A. Majoran case.} 
Consider first the Majorana case, Eqs. \refer{Eq:Maj-deg}. In order not to obscure the notation it is assumed that $k=1$ and $^{\rm fam}$ superscripts are omitted; thus the total number of LWs is $n=f$. On the assumptions stated in Sec. \ref{Sec:NotAndAssum}, it is clear that Eq. \refer{Eq:GapEq-0} has, in the sense of formal power series, $n$ different and non-vanishing solutions \kY
\eq{\label{Eq:m-a-Series}
m_{(a)}=m^{\rm tree}_{(a)}+\cO(\hbar)\,,
}
such that $\re(m_{(a)}) > 0$. Define 
\eq{\label{Eq:m-matr-Maj}
m=\diag(m_{(1)}\,,\ldots\,,m_{(n)})\,.
}
It will be shown (in Sec. \ref{Sec:Proof}) that the $\hat{\sD}_{{\cF}}(p)$ matrix in the full propagator of two Majorana fields, Eq. \refer{Eq:G2},  
has the following simple form
\eq{\label{Eq:D-as}
\hat{\sD}_{\cF}(p)
=
\hat{\zeta}\,
[p^2-m^2]^{-1}[\ds{p}+m]\,
\hat{\zeta}^{\,{\rmt}}
+\text{[non-pole part]}
\,,
}
where
\eqs{\label{Eq:Zeta-hat}
\hat{\zeta}\,&=&\zeta_L\, P_L\, + \,\zeta_R\, P_R\,,\nn\\
\hat{\zeta}^{\,\rmt}&=&\zeta_L^{\ \rmt} P_L + \zeta_R^{\ \rmt}  P_R\,,
}
matrices $\ze_{L,R}$ (as well as $m$)  
carry only flavor indices, 
while columns of $\zeta_{L}$ and $\zeta_{R}$ are given, respectively, by vectors $\zeta_{L[a]}$ and $\zeta_{R[a]}$ in the flavor space 
\eq{\nn
\zeta_X=
\Bigg[
\bigg[\zeta_{X[1]}\bigg]\,\, \cdots\,\,
\bigg[\zeta_{X[n]}\bigg]
\Bigg]\,,
\qquad X=L,R,
}
obtained in the following way.   
Let $\xi_{[a]}$ be an eigenvector of $\bM^2_L(m_{(a)}^2)$, Eq. \refer{Eq:bM2-LR}, corresponding to the eigenvalue $m_{(a)}^2$ \kF 
\koment{str.Z17}
\eq{\label{Eq:xi-Eig}
\bM^2_L(m_{(a)}^2)\,  \xi_{[a]} 
= 
m_{(a)}^2\,
\xi_{[a]}\,,
} 
and obeying the following normalization condition 
\eq{\label{Eq:norm-cond}
\xi_{[a]}^{\,\, \rmt} \, 
\sM_{L}(m_{(a)}^2)\, 
\xi_{[a]}
=m_{(a)}\,, 
}
then \koment{str.Z30} \kG 
\eq{\label{Eq:zeta-L-a}
\zeta_{L[a]}=\cN(a)\,\xi_{[a]}\,,
}
with a normalizing factor \kTT \koment{str.Z32}
\eq{\label{Eq:cN-final}
\cN(a)
=
\Big\{
1-
\frac{1}{m_{(a)}}
\xi_{[a]}^{\ \rmt}\,\sM_L(m_{(a)}^2)\,
\bM^2_{L}{}^{\prime}(m_{(a)}^2)\,\xi_{[a]}
\Big\}^{-1/2}
\,,
}
where 
$
\bM^2_L{}^{\prime}(s) 
\equiv 
\dd\bM^2_L(s)/\dd s\,,
$
and
\koment{str.Z16}
\eq{\label{Eq:zeta-R-a}
\zeta_{R[a]}
=
\frac{1}{m_{(a)}}
\sZ_R(m_{(a)}^2)^{-1}
\sM_L(m_{(a)}^2)\,
\zeta_{L[a]}\,.
}
(Note that, on the assumptions stated above, Eqs. \refer{Eq:xi-Eig}-\refer{Eq:norm-cond}, determine $\xi_{[a]}$ uniquely up to a sign; one could worry that the condition \refer{Eq:norm-cond} cannot be imposed since e.g.
$[1,\ \ i]\,[1,\ \ i]^{\rmt}=0$, 
however such a pathology is impossible at the tree-level, and thus it is impossible for formal power series.) 

Moreover it will be shown that, if Feynman integrals contributing to $\sZ_{L,R}(p^2)$ and $\sM_{L,R}(p^2)$ do not acquire imaginary parts in a left neighborhood \kUU $\sU_a\subset\bR$ of 
$p^2=(m^{\rm tree}_{(a)})^2$
\eq{\nn
\sU_a\equiv
\big\{p^2\in\bR|\ \   (m^{\rm tree}_{(a)})^2-\vep < p^2\leq(m^{\rm tree}_{(a)})^2 \big\}\,,
\qquad \vep>0\,,
}
so that the following reality conditions are satisfied \kW
\koment{str.34}
\eq{\label{Eq:RealityCond}
\sZ_R(s)=\sZ_L(s)^\star\,,\quad
\sM_R(s)=\sM_L(s)^\star\,,\quad
\forall_{s\in\sU_a}\,,
}
then all terms of a formal power series $m_{(a)}$, Eq. \refer{Eq:m-a-Series}, are real, \kDD and conditions \refer{Eq:xi-Eig}-\refer{Eq:zeta-R-a} imply that 
$\zeta_{R[a]}$ is the complex conjugation of $\zeta_{L[a]}$. \\

If, in particular, conditions \refer{Eq:RealityCond} are satisfied for all $a=1,\ldots,n$, then matrices appearing in Eq. \refer{Eq:Zeta-hat} obey $\zeta_{R}=\zeta_{L}{}^{\star}$ and Eq. \refer{Eq:D-as} has a simple interpretation: the Majorana field $\psi$ in, e.g., the $\anti{{\rm MS}}$ scheme of dimensional regularization can be expressed in terms of its on-shell scheme counterpart $\psi_{{\rm OS}}$ (see e.g. \cite{AOKI}) as follows \kPP \kZZ
\eq{\label{Eq:psi_MS-psi_OS}
\psi = \hat{\zeta}\, \psi_{{\rm OS}}.   
}

What if only some of the particles are stable? If $\im(m_{(a_S)})=0$, then one can introduce a free (interaction picture) Majorana field $\Psi^{\widetilde{a}_S}$ of mass $m_{(a_S)}$ with canonically normalized propagator and define (recall that ${\widetilde{b}}$
 is the ``total" index, cf. Eq. \refer{Eq:MajoField})
\eq{
\Psi^{\widetilde{b}}_{{\rm red}} = 
\sum_{  \widetilde{a}_S }
\big[\, \hat{\zeta}\, \big]^{\widetilde{b}}_{\ \widetilde{a}_S} 
\Psi^{\widetilde{a}_S}\,,
}
where the summation runs over all ``stable indices". Clearly,  $\Psi_{{\rm red}}$ is a free quantum field and Eq. \refer{Eq:D-as} implies that the  chronological propagator of $\Psi_{{\rm red}}$ reproduces the behavior of propagator in Eq. \refer{Eq:G2} about all poles located on the real axis.  
Thus, $\Psi_{{\rm red}}$ is the field that appears in the LSZ-reduced formula 
for the $S$-operator describing the transitions between stable states \cite{Bec} \kI 

\eq{\label{Eq:S}
S=\,\,:\!\exp(\Si)\!:\, \exp(i\, W[J])\bigg|_{J=0}\,,
}
with  
\eq{\label{Eq:Si}
\Si= 
-\int\,\volel{x}\,\Psi^{\widetilde{b}}_{{\rm red}}(x)
\int\,\volel{y}\,\Gamma_{\widetilde{b}\widetilde{c}}(x,y)\,
\derf{}{J_{\widetilde{c}}(y)}
\,,
}
where $\Gamma_{\widetilde{b}\widetilde{c}}(x,y)$ is the Fourier transform of \refer{Eq:Gamma2}, normal ordering in Eq. \refer{Eq:S} refers to free quantum fields $\Psi_{{\rm red}}$, \kJ
while the connected generating functional $W[J]$ is related through the Legendre transform to the (renormalized) 1PI effective action $\Gamma[\psi]$ 
\eq{\nn
\Gamma[\psi]=W[\cJ^\psi] - \cJ^\psi\!\cdot\!\psi\,,
\qquad \ 
\derf{W[J]}{J_{\widetilde{b}}(x)}\Bigg|_{J=\cJ^\psi}
=\psi^{\widetilde{b}}(x)\,,
}
(in the last three equations, $\psi$ and $\Psi$ represent not only fermions but also scalars and vectors).   

What about unstable particles? Consider, for instance, a theory in which heavy neutrinos described in terms of Majorana fields $\psi_{N}^{\bar{a}}$ carrying a family index $\bar{a}$, interact with a Hermitian scalar field $h$ and massless SM (anti)neutrinos, described in terms of Majorana fields $\psi_{\nu}^{{\check{b}}}$ carrying a family index $\check{b}$, through the following Lagrangian density (spinor indices are suppressed) 
\koment{str.PDCN24}
\eq{
\cL_{\rm int} = 
-
h\, \bar\psi_{N}^{\bar{a}}
\left(
\cY_{\bar{a} \check{b}}P_L+
\cY^{\,\star}_{\bar{a} \check{b}} P_R\right)
\psi_{\nu}^{\check{b}}\,.
}
The CP-asymmetry 
\eq{\label{Eq:ep-def-nowa}
\vep_{\bar{a} \check{b}}
=
\frac{\Gamma(N_{\bar{a}}\to h\nu_{\check{b}})-\Gamma(N_{\bar{a}}\to h\bar\nu_{\check{b}})}
     {\Gamma(N_{\bar{a}}\to h\nu_{\check{b}})+\Gamma(N_{\bar{a}}\to h\bar\nu_{\check{b}})}\,,
}
was calculated in \cite{PilafUnd,PilafRL1,PlumOld,PlumNew} by looking at diagrams in which $N_{\bar{a}}$ is an internal (rather than an external) line, what leads to the following expression \koment{PDCN20,PDCN17,PDCN12}
\eq{\label{Eq:ep-form}
\vep_{\bar{a} {\check{b}}}
=
\frac{|\cY^{R}_{\bar{a} {\check{b}}}|^2-|\cY^{L}_{\bar{a} {\check{b}}}|^2}
     {|\cY^{R}_{\bar{a} {\check{b}}}|^2+|\cY^{L}_{\bar{a} {\check{b}}}|^2}
\,,
}
with \koment{str.PDNC26}
\eqs{
\label{Eq:cY-L}
\cY^{L}_{\bar{a} {\check{b}}}
&=& \cY_{\bar{a}' {\check{b}}}(\zeta_L)^{\bar{a}'}_{\ \bar{a}}+\ldots\,,\\
\label{Eq:cY-R}
\cY^{R}_{\bar{a} {\check{b}}}
&=& \cY^{\,\star}_{\bar{a}' {\check{b}}}(\zeta_R)^{\bar{a}'}_{\ \bar{a}}+\ldots\,,
}
where 
$\zeta_{R,L}$ are $\zeta$ matrices for the $\psi_{N}^{\bar{a}}$ fields, while
the ellipsis indicates contributions of corrections to external lines of $h$ and $\psi_{\nu}^{{\check{b}}}$ fields, as well as loop corrections to the 1PI vertices (for simplicity it is assumed here that the mixing between light and heavy neutrinos is negligible, \kKKK even though the present formalism is capable of describing quantum corrections to the mixing in the full 6$\times$6 system). \kJJJ  \kBBB \kYYY
\\

\noindent {\bf \ref{Sec:PrescrFerm}.B Dirac case.} Consider now the Dirac case, Eqs. \refer{Eq:Dirac-Case}. For simplicity of the notation it is assumed that $\ell=1$, thus the total number of LWs is $n=2f$. On the assumptions stated in Sec. \ref{Sec:NotAndAssum}, it is clear that Eq. \refer{Eq:GapEq-0}, cf. Eqs. \refer{Eq:+Eq-}-\refer{Eq:sX-Dirac}, has $f$ different and non-vanishing solutions 
\eq{\label{Eq:m-a-Series-Dirac}
m_{(a)}=m^{\rm tree}_{(a)}+\cO(\hbar)\,,
}
such that $\re(m_{(a)}) > 0$. Define 
\eq{\label{Eq:m_D}
m_D=\diag(m_{(1)}\,,\ldots\,,m_{(f)})\,,
}
and 
\eq{\label{Eq:tm-Dirac}
\tm=
\left[
\begin{array}{cc}
 0   & m_D \\
 m_D &  0 \\
\end{array}
\right]\,.
}
It will be shown that the $\hat{\sD}_{{\cF}}(p)$ matrix in the full propagator of two Majorana fields, Eq. \refer{Eq:G2}, has the form  
\eq{ \label{Eq:D-as-Dirac}
\hat{\sD}_{\cF}(p)
=
\hat{\zeta}\,
[p^2-\tm^2]^{-1}[\ds{p}+\tm]\,
\hat{\zeta}^{\,{\rmt}}
+\text{[non-pole part]}
\,,
}
where
\eq{\label{Eq:Zeta-hat-Dirac}
\hat{\zeta}=\zeta_L\, P_L + \zeta_R\, P_R\,,
}
while the $\zeta_{L,R}$ matrices have a block-diagonal form 
\eq{\label{Eq:zeta_LR-Dirac}
\zeta_X=
\left[
\begin{array}{cc}
\bar\zeta_{X+}   & 0  \\
 0           & \bar\zeta_{X-}  \\
\end{array}
\right]\,,\qquad\quad X=L,R,
}
with matrices $\bar\zeta_{X\pm}$ built out of vectors $\bar\zeta_{X[a\pm]}$
\eq{\nn
\bar\zeta_{X\pm}
=
\Bigg[
\bigg[\bar\zeta_{X[1\pm]}\bigg]\, \cdots\,
\bigg[\bar\zeta_{X[f\pm]}\bigg]
\Bigg]\,,\qquad X=L,R,
}
obtained in the following way.    
Let $\bar\xi_{[a\pm]}$ be arbitrary \emph{but fixed} eigenvectors of $\bM^2_{\pm}(m_{(a)}^2)$, Eqs. \refer{Eq:Dirac-Case-bMpmkw}, corresponding to the eigenvalue $m_{(a)}^2$ \kN  
\koment{str.T$^3$}
\eqs{\label{Eq:xi-Eig-Dirac}
\bM^2_+(m_{(a)}^2)\,  \bar\xi_{[a+]} 
&=& 
m_{(a)}^2\,
\bar\xi_{[a+]}\,,
\nn\\
\bM^2_-(m_{(a)}^2)\,  \bar\xi_{[a-]} 
&=& 
m_{(a)}^2\,
\bar\xi_{[a-]}\,,
}
(eigenspaces of $\bM^2_{\pm}(m_{(a)}^2)$ are one-dimensional on the assumptions stated in Sec. \ref{Sec:NotAndAssum}),  
and obeying the following normalization condition 
\eq{\label{Eq:norm-cond-Dirac}
\bar\xi_{[a+]}^{\,\, \rmt} \, 
\mu_{L}(m_{(a)}^2)\, 
\bar\xi_{[a-]}
=m_{(a)}\,. 
}
Then \koment{str.C$^4$} \kG 
\eqs{\label{Eq:zeta-L-a-Dirac}
\bar\zeta_{L[a+]}&=&\cc(a)\,\bar\cN(a)\,\bar\xi_{[a+]}\,, \nn\\
\bar\zeta_{L[a-]}&=&\cc(a)^{-1}\,\bar\cN(a)\,\bar\xi_{[a-]}\,, 
}
and 
\eqs{\label{Eq:zeta-R-a-Dirac}
\bar\zeta_{R[a+]}
&=&
\frac{1}{m_{(a)}}\,
\sI_+(m_{(a)}^2)^\rmt\mu_L(m_{(a)}^2)\,
\bar\zeta_{L[a-]}\,,
\nn\\
\bar\zeta_{R[a-]}
&=&
\frac{1}{m_{(a)}}\,
\sI_-(m_{(a)}^2)^\rmt\mu_L(m_{(a)}^2)^\rmt\,
\bar\zeta_{L[a+]}\,,
}
with a normalizing factor \koment{str.B$^4$ $\&$ $I^4$}
\eq{\label{Eq:cN-final-Dirac}
\bar\cN(a)
=
\Big\{
1-
\frac{1}{m_{(a)}}
\bar\xi_{[a-]}^{\ \rmt}\,\mu_L(m_{(a)}^2)^\rmt\,
\bM^2_{+}{}^{\prime}(m_{(a)}^2)\,\bar\xi_{[a+]}
\Big\}^{-1/2}
\,,
}
where 
$
\bM^2_+{}^{\prime}(s) 
\equiv 
\dd\bM^2_+(s)/\dd s\,,
$
while $\cc(a)\in\bC\setminus\{0\}$ is an arbitrary number which does not affect the pole part of the propagator. 

Moreover it will be shown that, if the reality conditions \refer{Eq:RealityCond} hold in a left neighborhood $\sU_a\subset\bR$ of 
$p^2=(m^{\rm tree}_{(a)})^2$,
then all terms of a formal power series $m_{(a)}$, Eq. \refer{Eq:m-a-Series-Dirac}, are real, and there exists $\cc(a)\in\bC\setminus\{0\}$ such that
\eq{\label{Eq:aaa} 
\bar\zeta_{R[a+]}=\bar\zeta_{L[a+]}^{\ \ \star}\,, 
\qquad \text{and} \qquad
\bar\zeta_{R[a-]}=\bar\zeta_{L[a-]}^{\ \ \star}\,.
}
With fixed $\bar\xi_{[a\pm]}$, conditions \refer{Eq:aaa} determine $\cc(a)$ uniquely up to a phase. Thus, if conditions \refer{Eq:RealityCond} are satisfied for all $a=1,\ldots,f$, then matrices appearing in Eq. \refer{Eq:Zeta-hat-Dirac} obey $\zeta_{R}=\zeta_{L}{}^{\star}$. \kH \kNN\\

\noindent {\bf \ref{Sec:PrescrFerm}.C. Generic fermionic case.} 
The above prescriptions can be generalized to the case constrained only by the three conditions discussed below Eq. \refer{Eq:GapEq-0}.
Recall that these conditions imply that 
the number of poles of the full propagator is equal to the total number $n$ of LWs. The $a$ label is assumed to distinguish different solutions $m_{(a)}^2$ to  Eq. \refer{Eq:GapEq-0}; indices corresponding to this eigenvalue are labeled with $a_1,\ a_2,\ $ etc.
   
Let $\tm=\tm^\rmt$ be an \emph{arbitrarily} chosen symmetric $n\times n$ matrix such that \koment{str.Z8} 
\eq{\label{Eq:tm-sq}
\tm^2=\diag(m^2_{(1)}\,,\ldots)\,,
}
and  \koment{str.Z69}
\eq{\label{Eq:tm-zero}
\tm_{a_r b_q}=0\,\ \ \forall_{a_r}\,, \qquad \text{if} \qquad m_{(b)}^2=0\,.
}
Clearly,
\eq{\nn
[\tm^2]_{a_r a_q}= m_{(a)}^2\,\delta_{r q}\,. 
}
It will be shown that the $\hat{\sD}_{{\cF}}(p)$ matrix in the full propagator of two Majorana fields, Eq. \refer{Eq:G2},  
has the following form \koment{str.Z9}
\eq{\label{Eq:D-as-GENERAL}
\hat{\sD}_{\cF}(p)
=
\hat{\zeta}\,
[p^2-\tm^2]^{-1}[\ds{p}+\tm]\,
\hat{\zeta}^{\,{\rmt}}
+\text{[non-pole part]}
\,,
}
where
\eq{\nn 
\hat{\zeta}=\zeta_L\, P_L + \zeta_R\, P_R\,,
}
matrices $\ze_{L,R}$ (as well as $\tm$)  
carry only flavor indices, 
while columns of $\zeta_{L,R}$ are given by vectors $\zeta_{L,R[a_r]}$
\eq{\label{Eq:zeta-X-matrices}
\zeta_X=
\Bigg[
\bigg[\zeta_{X[1_1]}\bigg]\,\,\, \cdots\,\,
\Bigg]\,,
\qquad X=L,R,
}
(the order of columns reflects the order of eigenvalues in Eq. \refer{Eq:tm-sq}) 
obtained in the following way. 

\vspace*{5pt}
\noindent $\mathpzc{1}$. {\emph{Nonzero}} $m_\a^2$. 

\vspace*{5pt}
Let $\xi_{[a_1]}\,,\ldots,$ be a basis of the eigenspace \kF
\koment{str.Z17}
\eq{\label{Eq:xi-Eig-GENERAL}
\bM^2_L(m_{(a)}^2)\,  \xi_{[a_r]} 
= 
m_{(a)}^2\,
\xi_{[a_r]}\,,
}
obeying  the following normalization conditions
\eq{\label{Eq:norm-cond-GENERAL}
\xi_{[a_r]}^{\,\, \rmt} \, 
\sM_{L}(m_{(a)}^2)\, 
\xi_{[a_q]}
=\tm_{a_r a_q}\,, 
}
(recall that for each pair of nonsingular complex symmetric matrices $S_{1,2}$ there always exists a nonsingular matrix $N$ such that \kZ 
$S_{1}=N^\rmt\! S_2 N\,,$ thus starting with an accidentally chosen basis of eigenspace one can always find vectors obeying Eq. \refer{Eq:norm-cond-GENERAL}; the non-singularity of the left-hand side of Eq. \refer{Eq:norm-cond-GENERAL} is ensured by the assumptions listed below Eq. \refer{Eq:GapEq-0}\kMMM). 
Define the following matrix \koment{str.Z28}
\eq{\label{Eq:Xi}
\Xi(a)_{qr} = 
\xi_{[a_q]}^{\ \rmt}\,\sM_L(m_{(a)}^2)\,
\bM^2_{L}{}^{\prime}(m_{(a)}^2)\,\xi_{[a_r]}\,,
}
which is symmetric (see Sec. \ref{Sec:Proof}) and find a matrix $\cN(a)$ such that \koment{str.Z30}
\eq{\label{Eq:cN-cond}
\frac{1}{m_{(a)}^2}\,
\cN(a)\, \anti{m}(a)\, \cN(a)^\rmt
=
\big(\anti{m}(a)-\Xi(a)\big)^{-1}\,,
}
where \koment{str.Z27}
\eq{\label{Eq:bar-m(a)}
\anti{m}(a)_{rq}=\tm_{a_r a_q}\,.
}
(Clearly, $\cN(a)$ is determined only up to a complex orthogonal matrix.)   
Then \koment{str.Z30} \kG 
\eq{\label{Eq:zeta-L-a-GENERAL}
\zeta_{L[a_r]}=\sum_q\cN(a)^{q}_{\ r}\,\xi_{[a_q]}\,,
}
and
\koment{str.Z16/Z29}
\eq{\label{Eq:zeta-R-a-GENERAL}
\zeta_{R[a_r]}
=
\frac{1}{m^2_{(a)}}
\sZ_R(m_{(a)}^2)^{-1}
\sM_L(m_{(a)}^2)\,
\sum_q\zeta_{L[a_q]}\anti{m}(a)_{qr}\,.
}

Moreover it will be shown that, if  the $\tm$ matrix is chosen to be diagonal  \koment{str.Z61}
\eq{\nn
\tm=\diag(m_{(1)}\,,\ldots)\,,
}
with $\re (m_{(a)})>0$, and if reality conditions \refer{Eq:RealityCond} are satisfied in a left neighborhood $\sU_a\subset\bR$ of 
$p^2=(m^{\rm tree}_{(a)})^2$, then 
all terms of a formal power series $m_{(a)}$ are real and there exists a $\cN(a)$ matrix obeying Eq. \refer{Eq:cN-cond} and such that $\zeta_{R[a_r]}=\zeta_{L[a_r]}^{\ \star}$ for all $r$. With fixed $\{\xi_{[a_r]}\}$ eigenvectors this matrix is unique up to a \emph{real} orthogonal matrix $\cR(a)$, \ i.e. \  $\cN(a)=\cN_0(a)\cR(a)$.

\vspace*{5pt}

\noindent $\mathpzc{2}$. {\emph{Vanishing}} $m_\a^2$. 

\vspace*{5pt}

Let $\xi_{[0_1]}\,,\ldots,$ be a basis of the null eigenspace \kF
\eq{\label{Eq:xi-Eig-GENERAL-0}
\bM^2_L(0)\,  \xi_{[0_r]} 
= 0\,,
}
obeying the following normalization conditions \koment{str.Z81}
\eq{\label{Eq:norm-cond-GENERAL-0}
\xi_{[0_r]}^{\,\, \dagger} \, 
\sZ_{L}(0)\, 
\xi_{[0_q]}
=\de_{r q}\,, 
}
(for $p^2=0$ reality conditions \refer{Eq:RealityCond} cannot be violated and thus $\sZ_{L}(0)$ is a Hermitian and positive matrix\kVV, cf. Eqs. \refer{Eq:Z_LR-expansion} and \refer{Eq:AntiSym}). 
Define the following matrix \koment{str.Z80}
\eq{\label{Eq:Xi-0}
{\Xi}(0)_{qr} = 
\xi_{[0_q]}^{\ \dagger}\,\sZ_L(0)\,
\bM^2_{L}{}^{\prime}(0)\,\xi_{[0_r]}\,,
}
which is Hermitian (see Sec. \ref{Sec:Proof}) and find a matrix $\cN(0)$ such that \koment{str.Z83}
\eq{\label{Eq:cN-cond-0}
\cN(0)\, \cN(0)^\dagger=
\big(\id-\Xi(0)\big)^{-1}\,.
}
Then \koment{str.Z83} \kG 
\eq{\label{Eq:zeta-L-a-GENERAL-0}
\zeta_{L[0_r]}=\sum_q\cN(0)^{q}_{\ r}\,\xi_{[0_q]}\,,
}
and
\eq{\label{Eq:zeta-R-a-GENERAL-0}
\zeta_{R[0_r]}
=
\zeta_{L[0_r]}^{\ \star}
\,.
}\\

It should be stressed that auxiliary normalization conditions 
\refer{Eq:norm-cond-GENERAL} and \refer{Eq:norm-cond-GENERAL-0} are, in fact, redundant, i.e. prescriptions \refer{Eq:cN-cond} and \refer{Eq:cN-cond-0} for normalizing factors can be easily generalized to the case when the basis $\{\xi_{[a_r]}\}$ of eigenspace is completely arbitrary. \kWW Nonetheless, Eqs. 
\refer{Eq:norm-cond-GENERAL} and \refer{Eq:norm-cond-GENERAL-0} are imposed here, since the resulting equations \refer{Eq:cN-cond} and \refer{Eq:cN-cond-0} show immediately that, if flavor eigenfields are chosen to be canonically normalized eigenstates of the tree-level mass matrix, as is usually the case,  then the $\cN(a)$ matrix can be chosen as an $\cO(\hb)$ perturbation of the identity matrix, while $\xi_{[a_r]}$ can be chosen as $\cO(\hb)$ perturbations of vectors belonging to the canonical basis of $\bR^n\subset\bC^n$.\\

\noindent {\bf \ref{Sec:PrescrFerm}.D. Scalar case.}
Consider a set $\{\phi^\ell\}$ of $n$ scalar fields. 
Without loss of generality it is assumed that $\phi^\ell$ are Hermitian. 
The renormalized 1PI two-point function \koment{str.Y1}
\eqs{\label{Eq:Gamma2-scalar} 
\widetilde{\Ga}_{\ell j}(-p,p)
&=&
\Big[
p^2\id-(M^{\rm tree})^2-\Si(p^2)
\Big]_{\ell j}
\nn\\
&\equiv&
\Big[
p^2\id-M^2(p^2)
\Big]_{\ell j}
\,,
}
where $M^2(s)=M^2(s)^\rmt$ is a symmetric matrix, leads to the propagator
\eq{\label{Eq:G2-scalar} 
\widetilde{G}^{\,\ell j}(p,-p)
=
i 
\Big[ \big(p^2\id-M^2(p^2)\big)^{-1} \Big]^{\ell j}
\,,\
}
and the gap equation \kPPP
\eq{\label{Eq:GapEq-0-scalar}
\sX_S(m_{(\ell)}^2)=0\,,
}
with 
\eq{\label{Eq:sX_S}
\sX_S(s)\equiv\det(s\mathds{1}-M^2(s))\,.
}
It is assumed that assumptions listed below Eq. \refer{Eq:GapEq-0} for fermionic solutions $m_{(a)}^2$ and matrices $\bM^2_L(m_{(a)}^2)$, are satisfied also for their scalar counterparts, $m_{(\ell)}^2$ and $M^2(m_{(\ell)}^2)$. \kYY

Let $m^2$ be a diagonal $n\times n$ matrix  
\eq{\label{Eq:tm-sq-scalars}
m^2=\diag(m^2_{(1)}\,,\ldots)
\,.
}
The $\ell$ label is assumed to distinguish different values $m_{(\ell)}^2$; indices corresponding to this value in Eq. \refer{Eq:tm-sq-scalars} are labeled with $\ell_1,\ \ell_2,\ $ etc. 

It will be shown that the propagator \refer{Eq:G2-scalar} has the form
\eq{\label{Eq:D-as-GENERAL-scalars} 
\widetilde{G}(p,-p)
=
i\,\,  
{\zeta}\,
[p^2-m^2]^{-1}\,
{\zeta}^{\,{\rmt}}
+\text{[non-pole part]}
\,,
}
where columns of $\zeta$ are given by vectors $\zeta_{[\ell_r]}$
\eq{\label{Eq:zeta-X-matrices-scalars}
\zeta=
\Bigg[
\bigg[\zeta_{[1_1]}\bigg]\,\,\, \cdots\,\,
\Bigg]\,,
}
(the order of columns reflects the order of eigenvalues in Eq. \refer{Eq:tm-sq-scalars}) 
obtained in the following way.   
Let $\xi_{[\ell_1]}\,,\ldots,$ be a basis of the eigenspace \kQQ
\koment{str.Z17}
\eq{\label{Eq:xi-Eig-GENERAL-scalars}
M^2(m_{(\ell)}^2)\,  \xi_{[\ell_r]} 
= 
m_{(\ell)}^2\,
\xi_{[\ell_r]}\,,
}
obeying the following normalization conditions \koment{str.Y6}
\eq{\label{Eq:norm-cond-GENERAL-scalars}
\xi_{[\ell_r]}^{\,\, \rmt} \,  
\xi_{[\ell _q]}
=\delta_{r q}\,, 
}
(starting with an arbitrary basis of eigenspace one can always find vectors obeying Eq. \refer{Eq:norm-cond-GENERAL-scalars}, just as in the fermionic case). 
Define the following matrix \koment{str.Y10}
\eq{\label{Eq:Xi-scalars}
\Xi(\ell)_{qr} = 
\xi_{[\ell_q]}^{\ \rmt}\,
M^2{}^{\prime}(m_{(\ell)}^2)\,\xi_{[\ell_r]}\,,
}
which is manifestly symmetric, and find a matrix $\cN(\ell)$ such that \kSS \koment{str.Y9}
\eq{\label{Eq:cN-cond-scalars}
\cN(\ell)\, \cN(\ell)^\rmt
=
\big(\id-\Xi(\ell)\big)^{-1}\,.
}
(Clearly, $\cN(\ell)$ is determined only up to a complex orthogonal matrix.)   
Then \koment{str.Y4} \kG 
\eq{\label{Eq:zeta-L-a-GENERAL-scalars}
\zeta_{[\ell_r]}=\sum_q\cN(\ell)^{q}_{\ r}\,\xi_{[\ell_q]}\,.
}

Moreover it will be shown that, if Feynman integrals contributing to $M^2(p^2)$ do not acquire imaginary parts in a left neighborhood $\sU_\ell\subset\bR$ of 
$p^2=(m^{\rm tree}_{(\ell)})^2$, so that the following reality conditions are satisfied \koment{str.Y10 i nast.} \kRR
\eq{\label{Eq:RealityCond-scalar}
M^2(s)=M^2(s)^\star
\,,\qquad \ 
\forall_{s\in\sU_\ell}\,,
}
then all terms of a formal power series $m_{(\ell)}^2$ are real and there exists a $\cN(\ell)$ matrix obeying Eq. \refer{Eq:cN-cond-scalars} and such that $\zeta_{[\ell_r]}=\zeta_{[\ell_r]}^{\ \star}$ for all $r$. With fixed $\{\xi_{[\ell_r]}\}$ eigenvectors this matrix is unique up to a \emph{real} orthogonal matrix $\cR(\ell)$, \ i.e. \  $\cN(\ell)=\cN_0(\ell)\cR(\ell)$.\\

\noindent {\bf \ref{Sec:PrescrFerm}.E. Fermionic one-loop self-energy.} 
It is convenient to supplement the prescription for fermionic $\ze_{L,R}$ matrices by providing generic expressions for one-loop contributions in the $\anti{{\rm MS}}$ scheme with anticommuting $\gamma^5$ to the two-point functions $\sZ_{L,R}$ and $\sM_{L,R}$ in Eq. \refer{Eq:Gamma2}. 
Consider an arbitrary renormalizable model, in which Majorana fields $\psi^a$ 
(spinor indices are suppressed for simplicity) interact with Hermitian scalar fields $\phi^\ell$ (already shifted if necessary, so that $\VEV{\phi}=0$) and Hermitian gauge fields $A^\al_\mu$ via the following Lagrangian density \koment{por.(D:(D.15))i (D:(D.16))}
\eqs{\label{Eq:LagrTreeGI}
\mathcal{L}_{\rm int}^{\rm tree}&=&
+\frac{1}{2!}\,
i\,A^\alpha_\mu\,\,
\bar{\psi}^a\,\gamma^\mu 
\left(  \TF_{\alpha a b}\, P_L+\TF^\star_{\alpha a b}\, P_R \right)
\psi^b+\nn\\
&{}& 
-
\frac{1}{2!}\,\phi^\ell\,
\bar{\psi}^a 
\left(  \YF_{\ell a b}\, P_L+\YF^\star_{\ell a b}\, P_R \right)
\psi^b \,.
}
Here $\TF_{\alpha a b} = - \TF_{\alpha b a}^\star$ are matrix elements of ordinary anti-Hermitian gauge-group generators (already containing the coupling constants), while $\YF_{\ell a b}=\YF_{\ell b a}$ are matrix elements of symmetric Yukawa matrices. It is assumed that all fields are chosen to be the eigenfields of the tree-level mass-squared matrices, so that
\eqs{
\mathcal{L}_{\rm mass}^{\rm tree}&=&
+\frac{1}{2}\sum_{\beta} m_{V\!\beta}^2\,\eta^{\mu\nu} A_\mu^\beta A_\nu^\beta 
\ 
-\frac{1}{2}\sum_{\ell} m_{S\ell}^2\, \phi^\ell \phi^\ell 
+\nn\\
&{}&\nn
-\frac{1}{2}\,
\bar{\psi}^a 
\left(  M_{F a b}\, P_L+M^\star_{F a b}\, P_R \right)
\psi^b \,,
}
where \kNNN
\eq{\label{Eq:MF-kw-tree}
M_{F}M_{F}^\star = \diag(m_{F1}^2\,,\  m_{F2}^2\,,\ \ldots\,,\ m_{Fn}^2)\,,
} 
(clearly, without loss of generality one could assume that $M_{F}$ itself is diagonal;  such a choice is however completely impractical for Dirac particles, as it implies that, for instance, the $\cu$ matrix, Eq. \refer{Eq:cu}, in the SM is non-diagonal). 

Functions $\sZ_{L,R}$ and $\sM_{L,R}$ can be parametrized in the following way 
\eqs{
\sZ_{L,R}(p^2) &=&\id + \frac{\hb}{(4\pi)^2}\, \sZ_{L,R}^{(1)}(p^2) + \cO(\hb^2)\,,
\nn\\
\sM_{L}(p^2) &=& M_F + \hb\,\YF_{\ell}\,v_{(1)}^\ell+ \frac{\hb}{(4\pi)^2}\, \sM_{L}^{(1)}(p^2) + \cO(\hb^2)\,,
\nn\\
\sM_{R}(p^2) &=& 
         M_F^\star + \hb\,\YF_{\ell}^\star\,v_{(1)}^\ell+\frac{\hb}{(4\pi)^2}\, \sM_{R}^{(1)}(p^2) + \cO(\hb^2)\,,
}
where $v_{(1)}^\ell$ represents the one-loop contribution to the scalar vacuum expectation value (VEV), while $\sZ_{L,R}^{(1)}$ and $\sM_{L,R}^{(1)}$ 
are produced 
by one-loop diagrams shown in Figure 1 (we work in the Landau gauge). Using the standard, minimally subtracted one-loop functions $a^R$ and $b^R_0$ in the dimensional regularization \cite{Chank}
\eq{\nn 
a^R(m)
=
m^2\left\{\ln\frac{m^2}{\bar\mu^2}-1\right\}
\,,
\qquad\qquad\qquad\qquad\qquad\qquad
}

\vspace*{3 pt}
\eqs{\nn 
B_{M}\!\left(p^2,m_1,m_2\right)&\equiv&b_0^R\!\left(p^2,m_1,m_2\right)=
\\
&{}&\hspace*{-60 pt}
=
\int_0^1{\rm d}x\ln\frac{x(x-1)p^2+(1-x)m_1^2+x\,m_2^2-i\,0}{\bar\mu^2}
\,,\  
\nn
}
(here $\bar{\mu}$ is the renormalization scale of the $\anti{\rm MS}$ scheme, related to the usual \mbox{'t Hoot} mass unit via  
$\bar{\mu}\equiv\mu_H \sqrt{4\pi}\, e^{-{\gamma_E}/{2}}$), together with their 
combinations  $B_M(\equiv b^R_0)$, $B_Z$, $A_M$ and $A_Z$ \koment{str.C10,C8}
\eqs{\nn
B_{Z}\!\left(p^2,m_S,m_F\right)&=&
\frac{1}{2\,p^2}
\Big\{
a^{R}(m_F)-a^{R}(m_S) 
+
\qquad\qquad
\nn\\&{}&\hspace*{-50 pt}
+\left(m_S^2-m_F^2-p^2\right) b^{R}_0\!\left(p^2,m_S,m_F\right)
\Big\}
\nn\,,}

\vspace*{3 pt}
\eq{\nn
A_{M}(p^2,m_V,m_F) = 3\, b_0^R\!\left(p^2,m_V,m_F\right)+2\,,
\qquad\qquad\qquad\qquad\!\!\!\!\!\!\!\!\!\!\!\!\!
}

\vspace*{3 pt}
\eqs{\nn 
A_{Z}(p^2,m_V,m_F) 
&=&  
\frac{m_F^2+2 m_V^2-p^2}{2 p^2}\,\frac{a^R\!\left(m_V\right)}{m_V^2}+
\nn\\
&{}& \hspace*{-73 pt} 
+1-\frac{a^R\!\left(m_F\right)}{p^2}
+\frac{p^2+m_F^2-2 m_V^2}{2 p^2}\,\, b_0^R\!\left(p^2,m_V,m_F\right)
+\nn\\
&{}&\hspace*{-73 pt}
+\frac{\left(p^2-m_F^2\right){}^{\!2}}{2 p^2}\,  \frac{b_0^R\!\left(p^2,m_V,m_F\right)-b_0^R\!\left(p^2,0,m_F\right)}{m_V^2}\,,
\nn}
one gets \footnote{Correctness of these results was \emph{checked} with the aid of FeynCalc \cite{FC}.} \koment{str.C10,C11}
\eqs{
[\sZ_{L}^{(1)}(s)]_{ac} 
&=&\phantom{+}
\sum_{\be,b}A_{Z}(s\,,\, m_{V\!\be}\,,\, m_{Fb})\,\TF_{\be a b}\, \TF_{\be b c}
+\nn\\
&{}&+
\sum_{\ell,b}B_{Z}(s\,,\,m_{S\ell}\,,\,m_{Fb})\,
                                  \YF^\star_{\ell a b}\, \YF_{\ell b c}
\,,
\qquad\qquad
\nn
}

\eqs{
[\sZ_{R}^{(1)}(s)]_{ac} 
&=&\phantom{+}
\sum_{\be,b}A_{Z}(s\,,\, m_{V\!\be}\,,\, m_{Fb})\,
                           \TF_{\be a b}^\star\, \TF_{\be b c}^\star
+\nn\\
&{}&+
\sum_{\ell,b}B_{Z}(s\,,\,m_{S\ell}\,,\,m_{Fb})\,
                                  \YF_{\ell a b}\, \YF^\star_{\ell b c}
\,,\nn
\qquad\qquad
}

\eqs{
[\sM_{L}^{(1)}(s)]_{ac} 
&=&\!\!\phantom{+}
\sum_{\be,b,d}A_{M}(s\,,\, m_{V\!\be}\,,\, m_{Fb})\,
         \TF^\star_{\be a b}\,M_{F b d}\, \TF_{\be d c}
+\nn\\
&{}&\!\!+
\sum_{\ell,b,d}B_{M}(s\,,\,m_{S\ell}\,,\,m_{Fb})\,
                 \YF_{\ell a b}\,M_{F b d}^\star\, \YF_{\ell d c}
\,,\nn
}

\eqs{
[\sM_{R}^{(1)}(s)]_{ac} 
&=&\!\!\phantom{+}
\sum_{\be,b,d}A_{M}(s\,,\, m_{V\!\be}\,,\, m_{Fb})\,
         \TF_{\be a b}\,M_{F b d}^\star\, \TF_{\be d c}^\star
+\nn\\
&{}&\!\!+
\sum_{\ell,b,d}B_{M}(s\,,\,m_{S\ell}\,,\,m_{Fb})\,
                 \YF_{\ell a b}^\star\,M_{F b d}\, \YF_{\ell d c}^\star
\,.\nn
}
In particular, reality conditions \refer{Eq:RealityCond} are violated whenever $b_0^R$ has a non-vanishing imaginary part. 

Clearly, in the expression for $A_Z$,  the limits $m_{V\!\be}\to0$ are to be taken for contributions of massless gauge bosons. On the other hand, the last term in $A_{Z}$, even for spontaneously broken gauge symmetries, contains contributions of unphysical massless modes; as far as corrections to the pole masses are concerned,  they cancel with similar contributions of would-be Goldstone bosons, as the gauge symmetry leads to the following relation \kQQQ \koment{str.C14}
\eq{\label{Eq:goldi}
s_\ga\, m_{V\!\ga}\, \YF_{g}
=
M_{F}\,\TF_\ga - \TF_\ga^\star\,M_{F}
\,,
}  
where $\YF_{g}$ is a Yukawa matrix of the (massless) would-be Goldstone boson $\phi^g$ associated with a broken generator $\TF_\ga$, 
while $s_\ga=-1$ or $s_\ga=+1$. By contrast, contributions of unphysical modes to the $\zeta_{L,R}$ matrices do not cancel completely, but Eq. \refer{Eq:goldi} ensures that they do not contain resonant factors $(m_{Fa}^2-m_{Fb}^2)^{-1}$ \cite{PilafUnd}. Thus, in the CP-asymmetry \refer{Eq:ep-form}  these contributions cancel with similar ``unphysical" corrections to the 1PI vertices (indicated by the ellipsis in Eqs. \refer{Eq:cY-L}-\refer{Eq:cY-R}).  \kFFF

Finally, the one-loop contribution $v_{(1)}$ to the VEV can be obtained 
from the tadpole cancellation condition in the Landau gauge 
\eqs{\nn
0=
-\cV^{\prime }_{i}(v_{(0)}+\hb\,v_{(1)}) 
+
\frac{\hb}{(4\pi)^2}
\Big\{&{}&
3\sum_{\al j} [\TS_\al^2]_{ij}\,v_{(0)}^j
\left[a^R(m_{V\al})+\frac{2}{3}m_{V\al}^2\right]
+\nn\\[4mm]
&{}&\hspace*{-155 pt}
-\frac{1}{2}\sum_j \rho_{ijj}\,a^R(m_{Sj})
+\sum_{bc} ( M_{Fbc}^{\phantom{\star}} Y_{icb}^\star
             \!+\!M_{Fbc}^\star Y_{icb}^{\phantom{\star}}) 
            a^R(m_{Fb})
\Big\}
+\cO(\hb^2)\,.
\nn}
Here $\cV$ is the gauge-invariant tree-level potential of scalar fields, 
$v_{(0)}$ represents the tree-level VEV (i.e. $\cV^{\prime }_{i}(v_{(0)})=0$), 
$\rho_{ijk} = \mathcal{V}^{\prime\prime\prime}_{ijk}(v_{(0)})$, while $\TS_\al$ 
is the generator of the gauge group on scalar fields; $\TS_\al$ is normalized in such a way 
that the covariant derivative reads 
\eq{\nn
(D_\mu\phi)^j =\partial_\mu \phi^j
+ A^\alpha_\mu[{\TS}_\alpha]^{j}_{\ k}(\phi^k+v^k_{(0)}+\hb\,v^k_{(1)}+\ldots)\,.
}

\begin{figure}[t]
\hspace{0.0cm}
\centering
\includegraphics[scale=1.4]{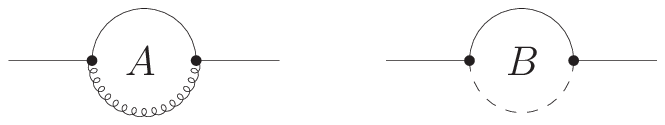}
\caption{One-loop contributions to $\sM_{L,R}$ and $\sZ_{L,R}$ in the Landau gauge. }
\end{figure}

\section{Proof} \label{Sec:Proof}

\noindent {\bf \ref{Sec:Proof}.A. Proof of generic fermionic prescription.} \\
The proof is a simple exercise in linear algebra. 

$\mathpzc{0}$. \emph{Generalities.} First of all, one has to calculate the following limits \kK (cf. Eqs. \refer{Eq:D-LR} and \refer{Eq:GapEq-0}) \koment{str.Z4}
\eq{\label{Eq:Delta-Def}
\Delta_{L,R}(a)=\!\!
\lim_{\ \, s\to m_{(a)}^2}
\Big\{
(s-m_{(a)}^2)\left[s\mathds{1}-\bM^2_{L,R}(s)\right]^{-1}
\Big\}
\,.
}
It is convenient to start with something simpler \koment{str.Z5}
\eq{ \label{Eq:bP(a)lim}
\bP(a)=\!\!
\lim_{\ \,  s\to m_{(a)}^2}
\Big\{
(s-m_{(a)}^2)\left[s\mathds{1}-\bM^2_{L}(m_{(a)}^2)\right]^{-1}
\Big\}
\,.
}
On the assumptions stated in Sec. \ref{Sec:NotAndAssum}, this limit exists and gives a projection onto the eigenspace of $\bM^2_{L}(m_{(a)}^2)$ associated with $m_{(a)}^2$ along the direct sum of remaining generalized eigenspaces of $\bM^2_{L}(m_{(a)}^2)$. \kT 
To verify this statement, it is enough to calculate the action of the right-hand side of Eq. \refer{Eq:bP(a)lim} on generalized eigenvectors of $\bM^2_{L}(m_{(a)}^2)$. \kHH
Introducing the resolvent \koment{str.Z47} 
\eq{\nn 
R(s)=\Big(s\id-\bM^2_{L}(m_{(a)}^2)\Big)^{-1}\,,
}
one has
\eq{\nn
R(s) \xi_{[a_r]} = (s-m_{(a)}^2)^{-1}\xi_{[a_r]}\,,
}
for all eigenvectors  $\xi_{[a_r]}$ associated with $m_\a^2$. Let $\la_\th\neq m_{(a)}^2$ be another eigenvalue of $\bM^2_{L}(m_{(a)}^2)$; the generalized eigenspace associated with it is spanned by (in general more than one) Jordan chain $\th_1\,,\ldots,\th_p$ (a subsequence of the Jordan basis for $\bM^2_{L}(m_{(a)}^2)$, see e.g. \cite{Axler}) \kU \koment{str.Z45}
\eq{\label{Eq:Jordan-chain}
\th_{r} = \big(\la_\th\id - \bM^2_{L}(m_{(a)}^2)\big)\th_{r+1}\,,
\qquad r=0,\ldots,p-1\,,
}
where $\th_0\equiv0$, i.e. $\th_1$ is an eigenvector. Let $Q(s)=(s-\la_\th)^{-1}$, the following identity can be easily checked by induction \kII \koment{str.Z49}
\eq{\label{Eq:R-th}
R(s)\,\th_r = \sum_{k=1}^r(-1)^{k+1}\,Q(s)^k\,\th_{r+1-k}\,.
}
Thus
\eq{\label{Eq:ker-im}
\bP(a)\,\xi_{[a_k]}=\xi_{[a_k]}\,,
\qquad
\bP(a)\,\th_r=0\,,
}
as was to be shown. In particular 
\eq{\label{Eq:bP(a)kw=bP(a)}
\bP(a)^2 = \bP(a) \,.
}

Expanding $\bM^2_{L}(s)$ in Eq. \refer{Eq:Delta-Def} about $s=m_{(a)}^2$ one gets \koment{str.Z7}
\eq{\label{Eq:Delta}
\Delta_{L}(a)=
\Big\{\mathds{1}-\bP(a)\,\bM^2_{L}{}^{\prime}(m_{(a)}^2)\Big\}^{-1}\,
\bP(a)\,.
}
Eq. \refer{Eq:bMRtrans} now yields 
\eq{\nn 
\Delta_{R}(a)
=
\sZ_R(m_{(a)}^2)^{-1}\,\Delta_{L}(a)^\rmt\,\!\sZ_R(m_{(a)}^2)\,.
}

Introducing another family $\{\tP(a)\}$  of projections
\eq{\nn
\tP(a)=\!\!
\lim_{\ \,  s\to m_{(a)}^2}
\Big\{
(s-m_{(a)}^2)\left[s\mathds{1}-\tm^2\right]^{-1}
\Big\}
\,,
}
one can decompose the $\tm^2$ 
matrix in Eq. \refer{Eq:tm-sq} as follows \koment{str.Z8}
\eq{\label{Eq:tm-na-rzuty}
\tm^2 = \sum_a m_{(a)}^2 \, \tP(a)\,,
}
clearly 
\eq{\nn
\sum_a \tP(a)=\mathds{1}\,, \qquad \text{and} \qquad \tP(a)\tP(b)=\delta_{a b}\tP(a)\,.
}
Now one sees that the formula that needs to be proven, Eq. \refer{Eq:D-as-GENERAL}, is equivalent to the following four sets of conditions 
($\bar s_a\equiv  m^2_\a$) \koment{str.Z10}

\eqs{
\label{Eq:Cond1}
\ze_{L}\,\tP(a)\,\ze_{R}^{\rmt}
&=&\De_L(a)\sZ_L(m_\a^2)^{-1}\,, \ \ \ \ \ \forall a,
\\
\label{Eq:Cond2}
\ze_{R}\,\tP(a)\,\ze_{L}^{\rmt}
&=&\sZ_R(m_\a^2)^{-1}\De_L(a)^{\rmt},\, \ \ \ \forall a,
}

\eqs{
\nn
\ze_{L}\,\tP(a)\,\tm\,\ze_{L}^{\rmt}
&=&\De_L(a)\sZ_L(\bar s_a)^{-1}\sM_R(\bar s_a)\sZ_R(\bar s_a)^{-1}\,,\ \ \forall a,
\\
\nn
\ze_{R}\,\tP(a)\,\tm\,\ze_{R}^{\rmt}
&=&\sZ_R(\bar s_a)^{-1}\De_L(a)^\rmt\!\sM_L(\bar s_a)\sZ_L(\bar s_a)^{-1}\,, \ \forall a.
}
Thus, one has to show  that there exist matrices $\zeta_{L,R}$ obeying, in addition to Eqs. \refer{Eq:Cond1}-\refer{Eq:Cond2}, the following conditions  
\eqs{
\nn
\ze_{L}\,\tP(a)\,\tm\,\ze_{L}^{\rmt}
&=&
\ze_{L}\,\tP(a)\,\ze_{R}^{\rmt}
\sM_R(m_\a^2)\sZ_R(m_\a^2)^{-1}\,,
\\
\nn
\ze_{R}\,\tP(a)\,\tm\,\ze_{R}^{\rmt}
&=&
\ze_{R}\,\tP(a)\,\ze_{L}^{\rmt}
\sM_L(m_\a^2)\sZ_L(m_\a^2)^{-1}\,.
}
It is enough \kS to impose, instead of the last two equations, the following two (cf. Eq. \refer{Eq:AntiSym}) \koment{str.Z12}
\eqs{
\label{Eq:Cond-3ini}
\ze_{R}\,\tm\,\tP(a)
&=&
\sZ_R(m_\a^2)^{-1}\sM_L(m_\a^2)\,\ze_{L}\,\tP(a)\,,
\\
\label{Eq:Cond-3iniprim}
\ze_{L}\,\tm\,\tP(a)
&=&
\sZ_L(m_\a^2)^{-1}\sM_R(m_\a^2)\,\ze_{R}\,\tP(a)
\,.
}
Using the following relation (cf. Eq. \refer{Eq:bar-m(a)}) \koment{str.Z68} 
\eq{\nn
\big[\tP(a)\,\tm\,\tP(a)\big]{}_{a_r a_q}=\anti{m}(a)_{rq}
\,,
}
together with the identity %
\footnote{This identity follows from 
$
(p^2-{\tm}^2)^{-1}\,\tm
=
\tm\, (p^2-\tm^2)^{-1}\,,
$
in the limit $p^2\to m_\a^2$.
}
$
\tm\,\tP(a)
=
\tP(a)\,\tm
\,,
$
which yields in turn\koment{str.Z68}
\eq{\nn
\tm\,\tP(a)
=
\tP(a)\,[\tP(a)\,\tm\,\tP(a)]\,,
} 
one can rewrite Eqs. \refer{Eq:Cond-3ini}-\refer{Eq:Cond-3iniprim} in terms of respective columns \kCCC of matrices $\zeta_{L,R}$ in Eq. \refer{Eq:zeta-X-matrices} \koment{str.Z68}

\eqs{ 
\label{Eq:Cond-3fin}
\sum_r \zeta_{R[a_r]}\, \anti{m}(a)_{rq}
&=&
\sZ_R(m_{(a)}^2)^{-1}
\sM_L(m_{(a)}^2)\,
\zeta_{L[a_q]}\,,\ \ \ \ \\
\label{Eq:Cond-3fin-prim}
\sum_r \zeta_{L[a_r]}\, \anti{m}(a)_{rq}
&=&
\sZ_L(m_{(a)}^2)^{-1}
\sM_R(m_{(a)}^2)\,
\zeta_{R[a_q]}\,.\ \ \ \ 
}

$\mathpzc{1}$. {\emph{Nonzero}} $m_\a^2$. 
Consider first the case $m_{(a)}^2\neq 0$; 
then Eq. \refer{Eq:Cond-3fin} is nothing more than the relation \refer{Eq:zeta-R-a-GENERAL}, since \kR 
\eq{\nn
\anti{m}(a)^2 = m_{(a)}^2\,\mathds{1}\,.
}
In turn, Eq. \refer{Eq:zeta-R-a-GENERAL} allows to rewrite Eq. \refer{Eq:Cond-3fin-prim} in an equivalent form \koment{str.Z14}
\eqs{\label{Eq:Cond4}
m_\a^2\,\zeta_{L[a_r]}
&=&\\\nn
&{}&
\hspace*{-55 pt}
=\sZ_L(m_\a^2)^{-1}\!\sM_R(m_\a^2)\sZ_R(m_\a^2)^{-1}\!\sM_L(m_\a^2)\zeta_{L[a_r]}
\,,
}
hence columns $\zeta_{L[a_1]},\ldots\,,$ of $\zeta_{L}$ are eigenvectors of $\bM_L^2(m_\a^2)$ corresponding to the eigenvalue $m_\a^2$, just as in Eqs. \refer{Eq:xi-Eig-GENERAL} and \refer{Eq:zeta-L-a-GENERAL}. 

Without loss of generality one can thus assume that, for $m_{(a)}^2\neq 0$,  $\zeta_{L[a_r]}$ are linear combinations of linearly independent eigenvectors  $\xi_{[a_q]}$ obeying the normalization conditions \refer{Eq:norm-cond-GENERAL}, with (yet unspecified) coefficients $\cN(a)^{q}_{\ r}$, as in Eq. \refer{Eq:zeta-L-a-GENERAL}.  It remains to be shown that  Eqs. \refer{Eq:Cond1}-\refer{Eq:Cond2} are equivalent to the condition  \refer{Eq:cN-cond} on the matrix $\cN(a)$. In fact Eq. \refer{Eq:Cond2}, being a transposition of \refer{Eq:Cond1}, can be skipped. Clearly, 
\eq{\nn
\ze_{L}\,\tP(a)\,\ze_{R}^{\rmt} 
= \sum_{q}  \ze_{L[a_q]}\ze_{R[a_q]}^{\ \rmt}\,.
}
Employing Eq. \refer{Eq:zeta-R-a-GENERAL}, 
and defining \koment{str.Z20,Z29}
\eq{\label{Eq:bY(a)def}
\bY(a) \equiv 
\sum_{q,r}\tau(i)^{qr}\,
\xi_{[a_q]} \, \xi_{[a_r]}^{\ \rmt} \, \sM_L(m_\a^2)\,,
}
with 
\eq{\label{Eq:tau}
\tau(a) \equiv
\frac{1}{m_{(a)}^2}\,
\cN(a)\, \anti{m}(a)\, \cN(a)^\rmt
\,,
}
one can rewrite Eq. \refer{Eq:Cond1} as \koment{str.Z21}
\eq{\nn
\bY(a)
=
\De_L(a)\,,
}
or, using Eq. \refer{Eq:Delta}, as
\eq{\label{Eq:Cond1-NEW}
\bY(a)
=
\Big\{\mathds{1}-\bP(a)\,\bM^2_{L}{}^{\prime}(m_{(a)}^2)\Big\}^{-1}\,
\bP(a)\,.
}
Eq. \refer{Eq:Cond1-NEW} can be further rewritten as 
\eq{
\label{Eq:Cond1-NEW2}
\bP(a)=\bS(a)\,,
}
where
\eq{\label{Eq:S(a)}
\bS(a) \equiv
\bY(a)
\Big\{\mathds{1}+\bM^2_{L}{}^{\prime}(m_{(a)}^2)\,\bY(a)\Big\}^{-1}
\,.
}
Note that the left-hand side of Eq. \refer{Eq:Cond1-NEW2} is a projection, cf. Eq. \refer{Eq:bP(a)kw=bP(a)}, thus it remains to be shown that $\cN(a)$ can be chosen in such a way that $\bS(a)$ is a projection operator with the same image and the same kernel as $\bP(a)$, cf. Eq. \refer{Eq:ker-im}. \kKK To that end, it is convenient to simplify first the explicit expression \refer{Eq:S(a)} for $\bS(a)$. Expanding a geometric series and appropriately changing the order of infinite sum with the summation over $q$ and $r$ appearing in Eq. \refer{Eq:bY(a)def} one ends up with another geometric series, thus \koment{Z24}
\eq{\label{Eq:S(a)simp}
\bS(a) = 
\sum_{q,r}
\Big[\Om(a)^{-1}\tau(a)\Big]^{q r}
\,
\xi_{[a_q]} \, \xi_{[a_r]}^{\ \rmt} \, \sM_L(m_\a^2)\,,
}
where \koment{str.Z24}
\eq{
\Om(a)\equiv\mathds{1}+\tau(a)\,\Xi(a)\,,
}
with the $\Xi(a)$ matrix defined in Eq. \refer{Eq:Xi}. 
The normalization condition for $\xi_{[a_r]}$ eigenvectors, Eq. \refer{Eq:norm-cond-GENERAL}, gives (cf. Eq. \refer{Eq:bar-m(a)}) \koment{str.Z25} 
\eq{\label{Eq:bS-xi}
\bS(a)\xi_{[a_r]}=\sum_q \big[\Om(a)^{-1}\tau(a)\anti{m}(a)\big]^{q}_{\ r}\,\,\xi_{[a_q]}\,,
}
hence \koment{str.Z25}
\eqs{\label{Eq:bS-kw}
\bS(a)^2
&=&
\sum_{q,r} 
\big[
\Om(a)^{-1}\tau(a)\anti{m}(a)
\Om(a)^{-1}\tau(a)
\big]^{qr}\, \times \nn\\
&{}&
\qquad\!\times\,\xi_{[a_q]} \, \xi_{[a_r]}^{\ \rmt} \, \sM_L(m_\a^2)\,.
}
Comparing this with Eq. \refer{Eq:S(a)simp} one sees that $\bS(a)$ is a projection operator if, for instance, the following equation is satisfied \koment{Z27}
\eq{\label{Eq:tau-Om-rel}
\tau(a)\anti{m}(a)=\Om(a)\,,
}
this is nothing more than the condition \refer{Eq:cN-cond}. 
To prove that a matrix $\cN(a)$ obeying Eq. \refer{Eq:cN-cond} indeed exists, it is necessary to show that the $\Xi(a)$ matrix, defined in Eq. \refer{Eq:Xi}, is symmetric. The following identity \koment{str.Z35} 
\eqs{\nn 
&{}&
\bM^2_{L}{}^{\prime}(s)^\rmt\sM_L(s)
-
\sM_L(s)\,\bM^2_{L}{}^{\prime}(s)
= 
\nn \\
&{}&\hspace*{40 pt}
=
\sM_L^\prime(s)\,\bM^2_{L}{}(s)
-
\bM^2_{L}{}(s)^\rmt\sM_L^\prime(s)
\,,
\nn
}
is easy to verify (cf. Eq. \refer{Eq:bM2-LR}); sandwiched between $\xi_{[a_q]}^{\ \rmt}$ and 
$\xi_{[a_r]}$ it gives 
\eq{
\Xi(a)_{rq}-\Xi(a)_{qr}=0\,,
}
since $\xi_{[a_{q,r}]}$ are eigenvectors of $\bM^2_{L}{}(m_{(a)}^2)$ corresponding to the same eigenvalue.

Moreover, Eqs. \refer{Eq:bS-xi} and \refer{Eq:tau-Om-rel} show that 
\eq{\label{Eq:bS-xi-fin}
\bS(a)\xi_{[a_r]} 
=
\xi_{[a_r]}
\,.
}
Hence, to complete the proof of the generalized prescription for $m_\a^2\neq0$, one has to show that the $\bS(a)$ operator annihilates these generalized eigenvectors 
of $\bM^2_{L}{}(m_{(a)}^2)$ which correspond to eigenvalues different than $m_{(a)}^2$, so that $\bS(a)=\bP(a)$. 
Because of Eq. \refer{Eq:S(a)simp} it is enough to prove the following property: let $\eta$ be an eigenvector of $\bM^2_{L}{}(m_{(a)}^2)$ corresponding to the eigenvalue $\la_{\eta}$ and let $\th$ be a generalized eigenvector of $\bM^2_{L}{}(m_{(a)}^2)$ associated with $\la_\th\neq\la_{\eta}$; then $\eta$ and $\th$ are $\sM_{L}{}(m_{(a)}^2)$-orthogonal
\eq{\label{Eq:orth}
\eta^{\rmt}\!\sM_{L}{}(m_{(a)}^2)\th=0\,.
}
This fact follows from the identity (cf. Eqs. \refer{Eq:bM2-LR} and \refer{Eq:AntiSym}) \koment{str.Z34 i Z52}
\eqs{\nn 
\bM^2_{L}{}(m_\a^2)^\rmt\sM_L(m_\a^2) 
-
\sM_L(m_\a^2)\,\bM^2_{L}{}(m_\a^2)
=0
\,.
\nn
}
Sandwiched between $\eta^{\rmt}$ and $\th_1$, i.e. the first element of a Jordan chain \refer{Eq:Jordan-chain}, it gives \koment{Z52, por. te¿ str.31/32}
\eq{\nn
(\la_\eta-\la_\th)\times
\eta^{\rmt}\!\sM_{L}{}(m_{(a)}^2)\th_1
=0
\,,
\nn
}
while sandwiched between $\eta^{\rmt}$ and $\th_{r+1}$ yields
\eq{\nn
(\la_\eta-\la_\th)\times
\eta^{\rmt}\!\sM_{L}{}(m_{(a)}^2)\th_{r+1}
=
-\eta^{\rmt}\!\sM_{L}{}(m_{(a)}^2)\th_{r}
\,.
\nn
}
This proves Eq. \refer{Eq:orth} by induction. \kLL

$\mathpzc{1}{\scriptstyle \frac{\mathpzc{1}}{\mathpzc{2}}}.$ \emph{Reality conditions.} Suppose now that conditions \refer{Eq:RealityCond} are satisfied for $s\in\sU_a\subset\bR$. 
Then $\sZ_L(s)$ is a Hermitian matrix, cf. Eq. \refer{Eq:AntiSym}, and thus one can parametrize it locally as \koment{str.37}
\eq{\nn 
\sZ_L(s)=U(s)^\dagger\,\La(s)\,U(s)\,,
}
where $U(s)$ is unitary, while $\La(s)$ is diagonal (and positive, cf. Eq. \refer{Eq:Z_LR-expansion}). 
On the other hand, a symmetric matrix 
\eq{\label{Eq:sM-tilde-par-stab}
{\widetilde{\sM}}_L(s)
\equiv 
\Big[\big(\sqrt{\La(s)}U(s)\big)^{-1}\Big]^{\rmt}
{{\sM}}_L(s)
\,\big(\sqrt{\La(s)}U(s)\big)^{-1}
\,,
\nn}
can be written in the following form
\eq{\nn
{\widetilde{\sM}}_L(s)
=V(s)^\rmt\mu(s)\,V(s)\,,
}
where $V(s)$ is unitary, while $\mu(s)$ is diagonal, real and nonnegative. 
Hence \koment{str.38}
\eq{\label{Eq:sZ-L-par}
\sZ_L(s)=\om(s)^\dagger \om(s)\,,
}
and
\eq{\label{Eq:sM-L-par}
\sM_L(s)=\om(s)^\rmt\mu(s)\om(s)\,,
}
where \koment{str.Z42}
\eq{
\omega(s)=V(s)\sqrt{\La(s)}U(s)\,.
}
Eq. \refer{Eq:bM2-LR} now reads
\eq{\label{Eq:bMstab-param}
\bM_L^2(s)=\om(s)^{-1}\mu^2(s)\om(s)\,,
}
where $\mu^2(s)\equiv \mu(s)^2$, and thus (cf. Eq. \refer{Eq:sX})
\eq{\label{Eq:sX-par}
\sX(s)=\prod_{\bar{c}}\Big(s-\mu_{\bar{c}\bar{c}}(s)^2\Big)\,.
}

Let $\{\bar{a}_r\}$ be a set of indices for which 
$\mu_{\bar{a}_r \bar{a}_r}(s)=m_\a^{\rm tree}+\cO(\hb)$. The gap equation \refer{Eq:GapEq-0} reduces to 
\eq{
  \mu_{\bar{a}_r \bar{a}_r}(m_{({\bar{a}_r})}^2) = m_{({\bar{a}_r})}\,.
}
A formal-power-series solution $m_{({\bar{a}_r})} = m_\a^{\rm tree}+\cO(\hb)$ to this equation obviously exists and is real, since all the derivatives 
$\mu_{\bar{a}_r \bar{a}_r}^{(k)}(s)$ at $s=(m^{\rm tree}_\a)^2$ are real. \kX
Let $\{{a}_r\}\subset\{\bar{a}_r\}$ be a set of indices for which $m_{({{a}_r})}=m_\a$; in other words, a situation in which the degeneracy of the tree-level masses is lifted by quantum corrections is not excluded here.   
Let $[\om(m_\a^2)^{-1}]_{[a_1]}\,,\ldots,$ be the columns of the $\om(m_\a^2)^{-1}$ matrix such that \koment{str.Z61}
\eq{
\mu_{a_{r} a_{r}}(m_\a^2)=m_\a\,,
}
clearly
\eq{\nn
[\om(m_\a^2)^{-1}]_{[a_r]}=\om(m_\a^2)^{-1}\,\mathds{1}_{[a_r]}\,.
}
Eigenvectors $\{\xi_{[a_u]}\}$, cf. Eq. \refer{Eq:xi-Eig-GENERAL}, have the form
\koment{Z62}
\eq{\nn 
\xi_{[a_u]}
=\sum_q C(a)^{q}_{\ u}\, \om(m_\a^2)^{-1} \id_{[a_q]}\,,
}
where $C(a)$ is a square matrix. 
The normalization condition \refer{Eq:norm-cond-GENERAL} reduces to (recall that $\tm$ is now assumed to be diagonal) \koment{Z62}
\eq{\label{Eq:CtrC=1}
C(a)^\rmt\,C(a)=\id\,,
}
i.e. $C(a)$ is a complex orthogonal matrix. 
The $\Xi(a)$ matrix, Eq. \refer{Eq:Xi}, reads \koment{Z62}
\eq{
\Xi(a)=C(a)^\rmt \Th(a)\, C(a)\,,
}
where  \koment{Z63}
\eq{\nn
\Th(a)_{ur}
=
m_\a\,\mathds{1}_{[a_u]}^\rmt\,\mu^{2\,\prime}(m_\a^2)\,\mathds{1}_{[a_r]}
=
m_\a\,\mu^{2\,\prime}_{a_u a_r}(m_\a^2)\,,
}
since terms with derivatives of $\omega(s)$ cancel. 
This shows that $\Th(a)$ is real. \kDDD (Since an accidental degeneracy of masses is not excluded, it is in principle possible that $\Th(a)$ is not proportional to the identity matrix.)

Eqs. \refer{Eq:zeta-L-a-GENERAL} and \refer{Eq:zeta-R-a-GENERAL} now read \koment{Z64/Z65}
\eqs{\label{Eq:xx1}
\zeta_{L[a_r]}&=&\sum_q \big[C(a)\cN(a)\big]^{q}_{\ r}\,
\big[\om(m_\a^2)^{-1}\big]\, \id_{[a_q]}\,,\\
\zeta_{R[a_r]}&=&\sum_q \big[C(a)\cN(a)\big]^{q}_{\ r}\,
\big[\om(m_\a^2)^{-1}\big]{}^{\!\star}\, \id_{[a_q]}\,
\,,
}
thus $\zeta_{R[a_r]}=\zeta_{L[a_r]}^{\ \star}$, if  $C(a)\cN(a)$ is a real matrix. Finally, with the aid of Eq. \refer{Eq:CtrC=1}, 
the condition \refer{Eq:cN-cond}  for $\cN(a)$  can be rewritten as \koment{Z64}
\eq{\label{Eq:xx3}
\big[C(a)\,\cN(a)]\,\big[C(a)\,\cN(a)]^{\rmt}
=
\Big\{\id-\frac{1}{m_{(a)}}\Th(a)\Big\}^{-1}\,.
}
Since the right-hand side of Eq. \refer{Eq:xx3} is a real diagonal and positive (in perturbation theory) matrix, there always exists a real matrix $C(a)\,\cN(a)$ obeying this condition. Clearly, Eq. \refer{Eq:xx3}, together with the reality condition $\zeta_{R[a_r]}=\zeta_{L[a_r]}^{\ \star}$,  determine $\cN(a)$ up to a rotation, as was to be shown.\\

$\mathpzc{2}$. {\emph{Vanishing}} $m_\a^2$. Consider the case $m_{(a)}^2=0$. Reality conditions \refer{Eq:RealityCond} cannot be violated for $p^2=0$, and thus Eqs. \refer{Eq:GapEq-0} and \refer{Eq:bM2-LR} give
\eq{\nn
\big| 
\det\big( \sM_{L}(0) \big)
\big|^2
=0=
\big| 
\det\big( \sM_{R}(0) \big)
\big|^2
.
}
Eqs. \refer{Eq:Cond-3fin}-\refer{Eq:Cond-3fin-prim} now show 
that columns $\zeta_{L,R[0_r]}$ have to belong to the kernel of $\sM_{L,R}(0)$, cf. Eq. \refer{Eq:tm-zero}, and therefore one can assume that Eq. \refer{Eq:zeta-R-a-GENERAL-0} holds. \kRRR One needs also the relation \koment{str.Z70}
\eq{\nn
\ker\bM^2_L(0) = \ker\sM_L(0) \,,
}
which follows immediately from the parametrization employed for analysis of reality conditions in the massive case, see Eqs.  \refer{Eq:sM-L-par} and \refer{Eq:bMstab-param}. 
Hence, one can assume that $\zeta_{L[0_r]}$ are linear combinations of linearly independent vectors $\xi_{[0_q]}$ obeying Eq. \refer{Eq:xi-Eig-GENERAL-0} and the normalization conditions \refer{Eq:norm-cond-GENERAL-0}, with (yet unspecified) coefficients $\cN(a)^{q}_{\ r}$, as in Eq. \refer{Eq:zeta-L-a-GENERAL-0}. 

It remains to be shown that  Eq. \refer{Eq:Cond1} reduces to the condition  \refer{Eq:cN-cond-0} on the matrix $\cN(0)$. This can be done just as before, by rewriting \refer{Eq:Cond1} as $\bP(0)=\bS(0)$ with $\bS(0)$ defined by \refer{Eq:S(a)} and appropriately adjusted matrix $\bY(0)$ \koment{str.Z71,por.Z80} 
\eq{\label{Eq:bY(a)def-0}
\bY(0) \equiv 
\sum_{q,r}\big[\cN(0)\,\cN(0)^\dagger\big]^{qr}\,
\xi_{[0_q]} \, \xi_{[0_r]}^{\ \dagger}\,\sZ_L(0)\,.
}
If Eq. \refer{Eq:cN-cond-0} is satisfied, one finds \koment{str.Z73,Z71,Z74,Z81} 
\eq{\label{Eq:bS(0)}
\bS(0) =
\sum_{q,r}\de^{qr}\,
\xi_{[0_q]} \, \xi_{[0_r]}^{\ \dagger}\,\sZ_L(0)\,,
}
and Eq. \refer{Eq:norm-cond-GENERAL-0} gives \koment{str.Z74,por.wz.Z.284}
\eq{\nn
{\rm im}\,\bS(0) \supset \ker\bM^2_L(0)\equiv {\rm im}\, \bP(0)\,.
}
The existence of matrices $\cN(0)$ obeying Eq. \refer{Eq:cN-cond-0} is ensured by the Hermiticity of $\Xi(0)$, which follows from the identity \kXX \koment{str.Z82} 
\eqs{\nn 
&{}&
\bM^2_{L}{}^{\prime}(0)^\dagger\sZ_L(0)^\dagger
-
\sZ_L(0)\,\bM^2_{L}{}^{\prime}(0)
= 
\nn \\
&{}&\hspace*{40 pt}
=
\sZ_L^\prime(0)\,\bM^2_{L}{}(0)
-
\bM^2_{L}{}(0)^\dagger\sZ_L^\prime(0)^\dagger
\,,
\nn
}
sandwiched between $\xi_{[0_q]}^{\ \dagger}$ and 
$\xi_{[0_r]}$. 

To complete the proof of $\bP(0)=\bS(0)$, one has to show that the generalized eigenvectors $\th$ of $\bM^2_L(0)$ associated with non-vanishing eigenvalues satisfy \koment{str.Z76}
\eq{\label{Eq:xxx}
\xi_{[0_r]}^{\ \dagger}\,\sZ_L(0)\,\th=0\,, \qquad  \forall_r\,.
} 
To that end one can employ once again the parametrization from  Eqs. \refer{Eq:sZ-L-par}-\refer{Eq:bMstab-param}. In particular, $\bM^2_L(0)$ is diagonalizable, and both $\xi_{[0_r]}$ and $\th$ are linear combinations of (disjoint sets of) columns of $\om(0)^{-1}$; hence Eq. \refer{Eq:xxx} follows immediately from Eq. \refer{Eq:sZ-L-par}. \\

\noindent {\bf \ref{Sec:Proof}.B. Proof of Majorana prescription.}  The prescription for Majorana case follows immediately from the generalized prescription, since one can take $\tm=m$, with a diagonal matrix $m$, Eq. \refer{Eq:m-matr-Maj}. In particular, the assumed non-degeneracy of the tree-level masses implies that $\cN(a)$  is a $1\times 1$ matrix and thus the freedom in Eq. \refer{Eq:cN-cond} reduces to a choice of sign. Hence, regardless of which sign is chosen, $\zeta_{R[a]}=\zeta_{L[a]}^{\ \star}$, if reality conditions  \refer{Eq:RealityCond} are satisfied for $s\in\sU_a\subset\bR$. \kEEE \\

\noindent {\bf \ref{Sec:Proof}.C. Proof of Dirac prescription.} Apart from Eq. \refer{Eq:aaa} 
for stable particles, the prescription for the Dirac case can be easily obtained from the generalized prescription. Having eigenvectors $\bar\xi_{[a\pm]}$, Eq. \refer{Eq:xi-Eig-Dirac}, it is convenient to choose eigenvectors $\xi_{[a_r]} $ (with 
$[a_r]=[a+],\,[a-]$), Eq. \refer{Eq:xi-Eig-GENERAL}, 
as \koment{str.T$^3$}
\eq{\label{Eq:xi-Dir}
\xi_{[a+]}=
\left[
\begin{array}{c}
\bar\xi_{[a+]}   \\
 0     
\end{array}
\right]\,,\qquad\quad 
\xi_{[a-]}=
\left[
\begin{array}{c}
0\\
\bar\xi_{[a-]} 
\end{array}
\right]\,,
}
and take $\cN(a)$ in Eq. \refer{Eq:zeta-L-a-GENERAL} to be the following 2$\times$2 matrix
\eq{
\cN(a)=
\left(\begin{array}{cc}
\bar{\cN}(a)\cc(a) & 0 \\
0 & \bar{\cN}(a)\cc(a)^{-1}
\end{array}
\right)
\,. 
} 
Similarly, a convenient choice of the $\tm$ matrix in Eq. \refer{Eq:tm-sq} is given by Eqs. \refer{Eq:tm-Dirac} and \refer{Eq:m_D}. With these choices, one-particle states corresponding to the columns of $\ze_{L,R}$ matrices, Eqs. \refer{Eq:zeta_LR-Dirac},  carry the  definite charge. \kBB The normalization condition \refer{Eq:norm-cond-GENERAL} now reduces to Eq. \refer{Eq:norm-cond-Dirac}, while Eq. \refer{Eq:cN-cond} is solved by \refer{Eq:cN-final-Dirac}.

Suppose that the reality conditions \refer{Eq:RealityCond} are satisfied for $s\in\sU_a\subset\bR$. Using the explicit form \refer{Eq:Dirac-Case} of $\sZ_{L,R}$ and $\sM_{L,R}$ matrices as well as the fact that an arbitrary nonsingular complex matrix $\widetilde{\mu_L}(s)$ can be written as \koment{str.E$^4$}
\eq{\label{Eq:mu-tilde-par-stab-DIRAC}
{\widetilde{\mu_L}}(s)
= 
V_+(s)^\rmt\mu(s)\,V_-(s)
\,,
\nn}
where $V_\pm(s)$ are unitary, while $\mu(s)$ is diagonal and positive, one finds (similarly as in the generic case) the following local parametrization of $\mu_{L}$ and $\sI_{\pm}^{-1}$ matrices  
\eqs{ \label{Eq:sZ-sM-L-par-DIRAC}
\sI_\pm(s)^{-1}&=&\om_\pm(s)^\dagger \om_\pm(s)\,, \nn\\
\mu_L(s)&=&\om_+(s)^\rmt\mu(s)\om_-(s)\,,
}
where
\eq{
\omega_{\pm}(s)=V_\pm(s)\La_\pm(s)U_\pm(s)\,,
}
with unitary $U_\pm(s)$, $V_\pm(s)$ matrices and positive-diagonal $\mu(s)$, $\La_\pm(s)$ matrices. 
Hence \koment{str.$F^4$}
\eq{\label{Eq:bMstab-param-DIRAC}
\bM_\pm^2(s)=\om_\pm(s)^{-1}\mu^2(s)\om_\pm(s)\,,
}
where $\mu^2(s)\equiv \mu(s)^2$, and thus (cf. Eq. \refer{Eq:+Eq-})
\eq{
\sX_+(s)=\prod_c\Big(s-\mu_{cc}(s)^2\Big)\,.
}
Thus, just like before, one sees that the solution $m_\a = m_\a^{\rm tree}+\cO(\hb)$ to the gap equation \refer{Eq:GapEq-0} is real and that the eigenvectors $\bar{\xi}_{[a\pm]}$ can be chosen as \koment{str.$H^4$}
\eq{\label{Eq:xi-Dir-choice-part}
\bar\xi_{[a\pm]}=[\om_\pm(m_\a^2)^{-1}]_{[a]}\,,
}
what ensures that the normalization condition \refer{Eq:norm-cond-Dirac} is obeyed. 
Eq. \refer{Eq:cN-final-Dirac}  now yields 
 \koment{str.$J^4$}
\eq{\nn\label{Eq:cN-stab-DIRAC}
\bar\cN(a)
=
\frac{1}{\sqrt
{
1-
\mu^2_{aa}{\!\!}^\prime(m_\a^2)
}}\,,
}
showing that $\bar\cN(a)$ is real. On the other hand, 
Eqs. \refer{Eq:zeta-R-a-Dirac} give \koment{str.$K^4$}
\eq{
\bar\zeta_{R[a\pm]}
=
{\bar\cN(a)}\,\cc(a)^{\mp 1}\,
\big(\bar\xi_{[a\pm]})^\star\nn
\,.
}
Comparing this with Eqs. \refer{Eq:zeta-L-a-Dirac} one sees that Eqs. \refer{Eq:aaa} hold provided that $\cc(a)$ is, for a \emph{particular} choice \refer{Eq:xi-Dir-choice-part}, a phase factor, as was to be proved.\\

\noindent {\bf \ref{Sec:Proof}.D. Proof of scalar prescription.} 
Similarly to the fermionic case one sees that Eq. \refer{Eq:D-as-GENERAL-scalars} is equivalent to the following conditions \kGGG \koment{str.Y2}
\eqs{
\label{Eq:CondS}
\ze\,\tP(\ell)\,\ze^{\rmt}
&=&\De(\ell)\,, \ \ \ \ \ \forall \ell,
}
where 
\eq{\nn
\tP(\ell)\equiv\!\!
\lim_{\ \,  s\to m_{(\ell)}^2}
\Big\{
(s-m_{(\ell)}^2)\left[s\mathds{1}-m^2\right]^{-1}
\Big\}
\,,
}
is a diagonal projection, while \koment{str.Y2}
\eqs{\label{Eq:Delta-Def-scalars}
\Delta(\ell)&\equiv&\!\!
\lim_{\ \, s\to m_{(\ell)}^2}
\Big\{
(s-m_{(\ell)}^2)\left[s\mathds{1}-M^2(s)\right]^{-1}
\Big\}\nn\\
&=&\ 
\Big\{\mathds{1}-\bP(\ell)\,M^2{}^{\prime}(m_{(\ell)}^2)\Big\}^{-1}\,
\bP(\ell)\,,
}
with $\bP(\ell)$ being the projection onto the eigenspace of $M^2(m_{(\ell)}^2)$ corresponding to $m_{(\ell)}^2$ along the direct sum of remaining generalized eigenspaces of $M^2(m_{(\ell)}^2)$. 

Clearly, \koment{str.Y3}
\eqs{
\ze\,\tP(\ell)\,\ze^{\rmt} 
&=& \sum_{q}  \ze_{[\ell_q]}\ze_{[\ell_q]}^{\ \rmt}
\,.
}
In the scalar case, there is no counterpart of  Eq. \refer{Eq:Cond4}; 
suppose then that $\zeta_{[\ell_r]}$ are linear combinations of (\emph{yet unspecified}) linearly independent vectors $\xi_{[\ell_q]}$ obeying the normalization condition \refer{Eq:norm-cond-GENERAL-scalars}, with (yet unspecified) coefficients $\cN(\ell)^{q}_{\ r}$, as in Eq. \refer{Eq:zeta-L-a-GENERAL-scalars}. 

Defining \koment{str.Y5}
\eq{\label{Eq:bY(a)def-scalars}
\bY(\ell) \equiv 
\sum_{q,r}\big[\cN(\ell)\,\cN(\ell)^\rmt\big]^{qr}\,
\xi_{[\ell_q]} \, \xi_{[\ell_r]}^{\ \rmt}\,,
}
one can (similarly to the fermionic case) rewrite Eq. \refer{Eq:CondS} as \koment{str.Y4}
\eq{
\label{Eq:Cond1-NEW2-scalars}
\bP(\ell)=\bS(\ell)\,,
}
where
\eqs{\label{Eq:S(a)-scalars}
\bS(\ell) 
&\equiv&
\bY(\ell)
\Big\{\mathds{1}+M^2{}^{\prime}(m_{(\ell)}^2)\,\bY(\ell)\Big\}^{-1}
\nn
\,,
}
what can be simplified to \koment{str.Y6}
\eq{\label{Eq:S(a)simp-scalars}
\bS(\ell) =
\sum_{q,r}
\big[\si(\ell)\big]^{q r}
\xi_{[\ell_q]} \, \xi_{[\ell_r]}^{\ \rmt}\,,
}
where \koment{str.Y6}
\eq{
\si(\ell)
\equiv
\Big\{\id+\cN(\ell)\,\cN(\ell)^\rmt\,\Xi(\ell)\Big\}^{-1}
\,\cN(\ell)\,\cN(\ell)^\rmt
\,,
}
with $\Xi(\ell)$ defined by Eq. \refer{Eq:Xi-scalars}.

The normalization condition for $\xi_{[\ell_r]}$ eigenvectors, Eq. \refer{Eq:norm-cond-GENERAL-scalars}, gives \koment{str.Y6} 
\eq{\label{Eq:bS-xi-scalars}
\bS(\ell)\xi_{[\ell_r]}=\sum_{q,s}
\big[\si(\ell)\big]^{qs}\,\de_{sr}\,\xi_{[\ell_q]}\,,
}
hence \koment{str.Y7}
\eqs{\label{Eq:bS-kw-scalars}
\bS(\ell)^2
&=&
\sum_{q,s,t,r} 
\si(\ell)^{qs}\,\de_{st}\,\si(\ell)^{tr}\,
 \xi_{[\ell_q]} \, \xi_{[\ell_r]}^{\ \rmt}\,.
}
Thus, the following condition (equivalent to Eq. \refer{Eq:cN-cond-scalars})
\eq{\label{Eq:tau-Om-rel-scalars}
\si(\ell)^{rs}=\de^{rs}\,,
}
ensures that $\bS(\ell)$ is a projection and that the image of $\bS(\ell)$ contains the subspace spanned by $\{\xi_{[\ell_q]}\}$. 
Therefore Eq. \refer{Eq:Cond1-NEW2-scalars} requires $\{\xi_{[\ell_q]}\}$ to be a basis of the eigenspace of $M^2(m_{(\ell)}^2)$ associated with $m_{(\ell)}^2$. To complete the proof of Eq. \refer{Eq:Cond1-NEW2-scalars}, one still has to show that the kernel of $\bS(\ell)$ is equal to the direct sum of generalized eigenspaces of $M^2(m_{(\ell)}^2)$ associated with eigenvalues different from $m_{(\ell)}^2$. This follows from the fact that a generalized eigenvector $\th$ of $M^2(m_{(\ell)}^2)$ associated with an eigenvalue $\la_\th$ is orthogonal to an eigenvector $\eta$ associated with $\la_\eta\neq\la_\th$ \koment{str.Y8}
\eq{\label{Eq:orth-scalars}
\eta^\rmt\,\th=0\,.
}
Eq. \refer{Eq:orth-scalars} can be proved in an analogous way to its fermionic counterpart, Eq. \refer{Eq:orth}, with the aid of the relation $M^2(s)^\rmt\equiv M^2(s)$. 

Suppose now that reality conditions \refer{Eq:RealityCond-scalar} are satisfied in a left neighborhood of 
$p^2=(m^{\rm tree}_{(\ell)})^2$. A real symmetric matrix $M^2(s)$ can be written as
\eq{\label{Eq:bMstab-param-scalars}
M^2(s)=\om(s)^{-1}\mu^2(s)\om(s)\,,
}
where $\mu^2(s)$ is diagonal and real,\kSSS while $\om(s)$ is a real orthogonal matrix. A similar argument to the one given below Eq. \refer{Eq:sX-par} shows that the pole mass squares are real.

Let $[\om(m_{(\ell)}^2)^{-1}]_{[\ell_1]}\,,\ldots,$ be the columns of the $\om(m_{(\ell)}^2)^{-1}$ matrix such that \koment{str.Y10}
\eq{
\mu^2_{\ell_{r} \ell_{r}}(m_{(\ell)}^2)=m_{(\ell)}^2\,.
}
Following the same reasoning as for the fermionic case in Sec. \ref{Sec:Proof}.A, one finds that columns (associated with $m_{(\ell)}^2$) of the $\zeta$ matrix, defined according to Eq. \refer{Eq:zeta-L-a-GENERAL-scalars},  have the form \koment{str.Y13} 
\eq{\label{Eq:xx1-scalars}
\zeta_{[\ell_r]}
=
\sum_q \big[C(\ell)\cN(\ell)\big]^{q}_{\ r}\,
\big[\om(m_{(\ell)}^2)^{-1}\big]\, \id_{[\ell_q]}\,,
}
where the $C(\ell)\cN(\ell)$ matrix obeys \koment{str.Y12}
\eq{\label{Eq:xx3-scalars}
\big[C(\ell)\,\cN(\ell)]\,\big[C(\ell)\,\cN(\ell)]^{\rmt}
=
\Big\{\id-\Th(\ell)\Big\}^{-1}\,,
}
with a real symmetric matrix $\Th(\ell)$ \koment{Y11}
\eq{\nn
\Th(\ell)_{qp}=
\id_{[\ell_q]}^{\, \rmt}\,
\om(m_{(\ell)}^2)\,
M^{2\,\prime}(m_{(\ell)}^2)\,
\om(m_{(\ell)}^2)^\rmt\,
\id_{[\ell_p]}
\,.
}
In particular, there exists a matrix $C(\ell)\,\cN(\ell)$ which obeys Eq. \refer{Eq:xx3-scalars} and is real, \kOO what ensures the reality of $\zeta_{[\ell_r]}$.

\section{Conclusions}\label{Sec:Concl}

We have analyzed in details the effects associated with mixing of scalar and 
fermionic fields. 
Presented results, together with their counterparts for vector fields  \cite{OnFeRu2},  
can be useful in the study of extensions of the Standard 
Model. In particular, the prescription for ``square-rooted residues" 
$\zeta$ is formulated entirely in terms of eigenvectors of certain matrices, and 
thus it 
can be efficiently employed in numerical calculations. 

\vspace{0.2cm}
\noindent{\bf Acknowledgments:} 
I am grateful to \mbox{P. Chankowski} for inspiring conversations and critical reading of an early version of this paper. I also thank to \mbox{K. Meissner} and \mbox{H. Nicolai} for helpful discussions.


\begin{thebibliography}{99}

\bibitem{MSbar}
  W.~A.~Bardeen, A.~J.~Buras, D.~W.~Duke and T.~Muta,
  Phys.\ Rev.\ D {\bf 18} (1978) 3998.
  doi:10.1103/PhysRevD.18.3998

\bibitem{Z-J:74} J.~Zinn-Justin, in
  \emph{Trends in Elementary Particle Theory:  International Summer
  Institute on Theoretical Physics in Bonn 1974}
  (Springer-Verlag, Berlin, 1975).


\bibitem{BRS1} C. Becchi, A. Rouet and R. Stora,
   Commun. Math. Phys. {\bf 42} (1975) 127-162.

\bibitem{BRS2}  C. Becchi, A. Rouet and R. Stora, 
   Annals of Phys. 98 (1976) 287-321.

\bibitem{PiguetSorella} O. Piguet and S. P. Sorella,
   \emph{Algebraic Renormalization}, Springer-Verlag Berlin Heidelberg  (1995).

\bibitem{Bon} G. Bonneau, Int. J. Mod. Phys. A{\bf 5} (1990) 3831-3860.

\bibitem{ChLM}
  P.~H.~Chankowski, A.~Lewandowski and K.~A.~Meissner,
  JHEP {\bf 1611} (2016) 105
  doi:10.1007/JHEP11(2016)105
  [arXiv:1608.02270 [hep-ph]].


\bibitem{Stapp}
  H.~P.~Stapp,
  Nuovo Cimento {\bf 32} (1964) 103.   

\bibitem{Stuart}
  R.~G.~Stuart,
  Phys.\ Rev.\ Lett.\  {\bf 70} (1993) 3193.
  doi:10.1103/PhysRevLett.70.3193

\bibitem{PilafRL1}
  A.~Pilaftsis,
  Phys.\ Rev.\ D {\bf 56} (1997) 5431
  doi:10.1103/PhysRevD.56.5431
  [hep-ph/9707235].

\bibitem{PlumOld}
  W.~Buchmuller and M.~Plumacher,
  Phys.\ Lett.\ B {\bf 431} (1998) 354
  doi:10.1016/S0370-2693(97)01548-7
  [hep-ph/9710460].

\bibitem{PilafUnd}
  A.~Pilaftsis and T.~E.~J.~Underwood,
  Nucl.\ Phys.\ B {\bf 692} (2004) 303
  doi:10.1016/j.nuclphysb.2004.05.029
  [hep-ph/0309342].


\bibitem{PlumNew}
  A.~Anisimov, A.~Broncano and M.~Plumacher,
  Nucl.\ Phys.\ B {\bf 737} (2006) 176
  doi:10.1016/j.nuclphysb.2006.01.003
  [hep-ph/0511248].

\bibitem{AOKI} 
  K.~I.~Aoki, Z.~Hioki, M.~Konuma, R.~Kawabe and T.~Muta,
  Prog.\ Theor.\ Phys.\ Suppl.\  {\bf 73} (1982) 1.
  doi:10.1143/PTPS.73.1


\bibitem{Velt} 
  M.~J.~G.~Veltman,
  Physica {\bf 29} (1963) 186.
  doi:10.1016/S0031-8914(63)80277-3 


\bibitem{Bec} 
  C.~Becchi,
  hep-th/9607181.


\bibitem{Fuchs}
  E.~Fuchs and G.~Weiglein,
  JHEP {\bf 1709} (2017) 079
  doi:10.1007/JHEP09(2017)079
  [arXiv:1610.06193 [hep-ph]].



\bibitem{Kniehl-1}
  B.~A.~Kniehl,
  Phys.\ Rev.\ D {\bf 89} (2014) no.9,  096005.
  doi:10.1103/PhysRevD.89.096005

\bibitem{Kniehl-2}
  B.~A.~Kniehl,
  Phys.\ Rev.\ Lett.\  {\bf 112} (2014) no.7,  071603
  doi:10.1103/PhysRevLett.112.071603
  [arXiv:1308.3140 [hep-ph]].

\bibitem{Kniehl-3}
  B.~A.~Kniehl,
  Phys.\ Rev.\ D {\bf 89} (2014) no.11,  116010
  doi:10.1103/PhysRevD.89.116010
  [arXiv:1404.5908 [hep-th]].

\bibitem{OnFeRu2}
  A.~Lewandowski,
``LSZ-reduction, resonances and non-diagonal propagators: 
gauge fields".

\bibitem{Weinberg} 
  S.~Weinberg,
  Phys.\ Rev.\ D {\bf 7} (1973) 1068.
  doi:10.1103/PhysRevD.7.1068


\bibitem{Gross-Wilczek} 
  D.~J.~Gross and F.~Wilczek,
  Phys.\ Rev.\ D {\bf 8} (1973) 3633.
  doi:10.1103/PhysRevD.8.3633



\bibitem{Jack-Osborn}
I. Jack and H. Osborn, Nucl. Phys. B249 (1985) 472.


\bibitem{Mach-Vaughn}
 E. Machacek and M. T. Vaughn, Nucl. Phys. B{\bf 222} (1983) 83,
  B{\bf 236} (1984) 221, B{\bf 249} (1985) 70.



\bibitem{Luo}
  M.~Luo, H.~Wang and Y.~Xiao,
  Phys.\ Rev.\ D {\bf 67} (2003) 065019
  doi:10.1103/PhysRevD.67.065019
  [hep-ph/0211440].



\bibitem{Martin} S.~P.~Martin,
  Phys.\ Rev.\ D{\bf 65} (2002) 116003  [hep-ph/0111209].

\bibitem{Martin-3}
  S.~P.~Martin,
  Phys.\ Rev.\ D {\bf 96} (2017) no.9,  096005
  doi:10.1103/PhysRevD.96.096005
  [arXiv:1709.02397 [hep-ph]].




\bibitem{Axler} S. Axler, 
\emph{Linear Algebra Done Right, 3rd edition}, Springer International Publishing, 2015.


\bibitem{Kibble}
  T.~W.~B.~Kibble,
  Phys.\ Rev.\  {\bf 174} (1968) 1882.
  doi:10.1103/PhysRev.174.1882



\bibitem{WeinT2} 
S.~Weinberg, {\it The quantum theory of fields}, Vol. II,
    Cambridge University Press, 1996.



\bibitem{Chank} P. H. Chankowski, \emph{Lectures on Quantum Field Theory},
  (unpublished).




\bibitem{FC}
  V.~Shtabovenko, R.~Mertig and F.~Orellana,
  Comput.\ Phys.\ Commun.\  {\bf 207} (2016) 432
  doi:10.1016/j.cpc.2016.06.008
  [arXiv:1601.01167 [hep-ph]].



\bibitem{NAKANISHI} 
  N.~Nakanishi,
  Prog.\ Theor.\ Phys.\ Suppl.\  {\bf 51} (1972) 1.
  doi:10.1143/PTPS.51.1









\bibitem{WeinT3} 
S.~Weinberg, {\it The quantum theory of fields}, Vol. III,
    Cambridge University Press, 2000.




 
\end{thebibliography}
\end{document}